\documentclass[journal]{IEEEtran}
\usepackage[noadjust]{cite}
\usepackage{tikzsymbols}
\usepackage{graphicx}
\usepackage{changepage}
\graphicspath{ {./images/} }
\usepackage{array}
\ifCLASSINFOpdf
\else
\fi
\usepackage[nolist]{acronym}

\begin{acronym}
\acro{ecu}[ECU]{Electronic Control Unit}
\acro{iot}[IoT]{Internet of Things}
\acro{iov}[IoV]{Internet of Vehicles}
\acro{cav}[CAV]{Connected Autonomous Vehicle}
\acro{cavs}[CAVs]{Connected Autonomous Vehicles}
\acro{dos}[DoS]{Denial of Service}
\acro{sql}[SQL]{Structured Query Language}
\acro{sae}[SAE]{Society of Automotive Engineers}
\acro{dgps}[DGPS]{Differential Global Positioning System}
\acro{slam}[SLAM]{Simultaneous Localization and Mapping}
\acro{oem}[OEMs]{Original Equipment Manufacturers}
\acro{lidar}[LIDAR]{Light Detection and Ranging}
\acro{imu}[IMU]{Inertial Measurement Unit}
\acro{rtk}[RTK]{Real Time Kinematic}
\acro{gnss}[GNSS]{Global Navigation Satellite System}
\acro{eea}[E/ E-Architecture]{Electrical/Electronic-Architecture}
\acro{tpms}[TPMS]{Tire Pressure Monitoring System}
\acro{ecu}[ECU]{Electronic Control Unit}
\acro{agps}[A-GPS]{Assisted GPS}
\acro{gsm}[GSM]{Global System for Mobile Communication}
\acro{gps}[GPS]{Global Positioning System}
\acro{isp}[ISP]{Internet Service Provider}
\acro{ssid}[SSID]{Service Set Identifier}
\acro{ttff}[TTFF]{Time To First Fix}
\acro{nrtk}[NRTK]{Network Real Time Kinematic}
\acro{hmi}[HMI]{Human-Machine Interface}
\acro{adas}[ADAS]{Advanced Driver Assistance System}
\acro{ml}[ML]{Machine Learning}
\acro{roi}[ROI]{Region of Interest}
\acro{ndt}[NDT]{Normal Distributions Transform}
\acro{autosar}[AUTOSAR]{AUTomotive Open System ARchitecture}
\acro{api}[API]{Application Programming Interface}
\acro{apis}[APIs]{Application Programming Interfaces}
\acro{os}[OS]{Operating System}
\acro{ara}[ARA]{AUTOSAR Runtime Environment for Adaptive Applications}
\acro{obu}[OBU]{On-Board Diagnostics}
\acro{eu}[EU]{European Union}
\acro{etsi}[ETSI]{European Telecommunications Standards Institute}
\acro{siem}[SIEM]{Security Information and Event Management}
\acro{mttc}[MTTC]{Mean Time-to-Compromise}
\acro{cvss}[CVSS]{Common Vulnerability Scoring System}
\acro{dread}[DREAD]{Damage Reproducibility Exploitability Affected users Discoverability}
\acro{stride}[STRIDE]{Spoofing, Tampering, Repudiation, Information Disclosure, Denial of Service, Elevation of Privilege}
\acro{dlt}[DLT]{Distributed Ledger Technology}
\acro{gdpr}[GDPR]{General Data Protection Regulation}
\acro{ttc}[TTC]{Time-to-Compromise}
\acro{asil}[ASIL]{Automotive Safety Integrity Level}
\acro{uml}[UML]{Unified Modeling Language}
\acro{it}[IT]{Information Technology}
\acro{cia}[CIA]{Confidentiality Integrity Availability}
\acro{capec}[CAPEC]{Common Attack Patterns Enumeration and Classification}
\acro{tal}[TAL]{Threat Agent Library}
\acro{octave}[OCTAVE]{Operationally Critical Threat, Asset, and Vulnerability Evaluation}
\acro{tvra}[TVRA]{Threat Vulnerability and Risk Analysis}
\acro{tara}[TARA]{Threat Agent Risk Assessment}
\acro{dfd}[DFD]{Data Flow Diagrams}
\acro{v2i}[V2I]{Vehicle to Infrastructure}
\acro{p2i}[P2I]{Peripheral to Infrastructure}
\acro{can}[CAN]{Controller Area Network}
\acro{iso26262}[ISO 26262]{Road Vehicles Functional Safety}
\acro{saej3061}[SAE J3061]{Cybersecurity Guidebook for Cyber-Physical Vehicle Systems}
\acro{v2c}[V2C]{Vehicle to Cloud}
\acro{mol}[MOL]{Methods and Objectives Library}
\acro{cel}[CEL]{Common Exposure Library}
\acro{risos}[RISOS]{Research in Secured Operating Systems}
\acro{pa}[PA]{Protection Analysis}
\acro{crud}[CRUD]{Create, Read, Update, and Delete}
\acro{ctc}[CTC]{Cost-to-Compromise}
\acro{poc}[PoC]{Proof of Concept}
\acro{ncva}[NCVA]{Network Communications Vulnerability Assessment}
\acro{osi}[OSI]{Open Systems Interconnection}
\acro{tpms}[TPMS]{Tire Pressure Monitoring System}
\acro{tcm}[TCM]{Telematics Control Module}
\acro{obd}[OBD]{On-Board Diagnostics}
\acro{hsm}[HSM]{Hardware Security Module}
\acro{mitm}[MITM]{man-in-the-middle}
\acro{bs}[BS]{Base Score}
\acro{ts}[TS]{Temporal Score}
\acro{es}[ES]{Environmental Score}
\acro{os}[OS]{Operating System}
\acro{ib}[IB]{Impact Bias}
\acro{cib}[CIB]{Confidentiality Impact Bias}
\acro{iib}[IIB]{Integrity Impact Bias}
\acro{aib}[AIB]{Availability Impact Bias}
\acro{av}[AV]{Access Vector}
\acro{ac}[AC]{Access Complexity}
\acro{a}[A]{Authentication}
\end{acronym}

\usepackage{graphicx}
\usepackage{circuitikz}

\usepackage{tikz}
\usetikzlibrary{decorations.pathreplacing}
\usetikzlibrary{shapes.geometric}
\usetikzlibrary{arrows.meta,calc,decorations.markings,math,arrows.meta}
\usetikzlibrary{positioning,shapes,backgrounds,matrix,calc,shadings,fit, optics}
\usetikzlibrary{automata,arrows}

\definecolor{orcidlogocol}{HTML}{A6CE39}
\tikzset{
    orcidlogo/.pic={
    \fill[orcidlogocol] svg{M256,128c0,70.7-57.3,128-128,128C57.3,256,0,198.7,0,128C0,57.3,57.3,0,128,0C198.7,0,256,57.3,256,128z};
    \fill[white] svg{M86.3,186.2H70.9V79.1h15.4v48.4V186.2z}
                 svg{M108.9,79.1h41.6c39.6,0,57,28.3,57,53.6c0,27.5-21.5,53.6-56.8,53.6h-41.8V79.1z M124.3,172.4h24.5c34.9,0,42.9-26.5,42.9-39.7c0-21.5-13.7-39.7-43.7-39.7h-23.7V172.4z}
                 svg{M88.7,56.8c0,5.5-4.5,10.1-10.1,10.1c-5.6,0-10.1-4.6-10.1-10.1c0-5.6,4.5-10.1,10.1-10.1C84.2,46.7,88.7,51.3,88.7,56.8z};
  },
    pics/person/.style n args={1}{
        code={
            \draw[thick,fill=black,transform shape = true] (0,0.925) circle(4pt);
            \draw[black,line width=8pt,transform shape = true] (0,0.75) -- ( 0,0.25);
            \draw[black,line width=3pt,transform shape = true] (-0.08,0.23) -- ( -0.08,-0.25);
            \draw[black,line width=3pt,transform shape = true] (0.08,0.23) -- ( 0.08,-0.25);
            \draw[black,line width=2pt,transform shape = true] (-0.2,0.75) -- ( -0.2,0.25);
            \draw[black,line width=2pt,transform shape = true] (0.2,0.75) -- ( 0.2,0.25);
        },
      },
    pics/parking/.style n args={1}{
        code = { %
            \draw (0,0) node[rectangle,fill={rgb:red,1;green,2;blue,5}, text=white, transform shape = true] {\Large \textbf{P}};
        }
    },
    myradiation/.style={{decorate,decoration={expanding waves,angle=90,segment length=4pt}}},
    pics/mybeacon/.style n args={1}{
        code = { %
            \draw[thick,fill=gray] (0,0) circle(20pt);
            \draw[thick] (0,0) circle(4pt);
            \draw[semithick,myradiation,decoration={angle=45}] (0cm+4pt,0) -- +(0:0.5);
            \draw[semithick,myradiation,decoration={angle=45}] (0cm-4pt,0) -- +(180:0.5);
        }
    },
    pics/antenna/.style n args={1}{
        code = { %
            \draw[black,line width=1pt,transform shape = true] (0,0) -- ( 0,-1);
            \draw[thick, fill=black] (0,0) circle(4pt);
            \draw[semithick,myradiation,decoration={angle=45}] (0cm+4pt,0) -- +(0:1);
            \draw[semithick,myradiation,decoration={angle=45}] (0cm-4pt,0) -- +(180:1);
        }
    },
    pics/cameraOne/.style n args={2}{
        code = { %
            \draw (0,0) node[draw,fill=#1, thick, minimum width=1cm,minimum height=2cm,transform shape = true] {};
            \draw (0.625,0.025) node[draw,fill=#2, thick, minimum width=0.2cm,minimum height=1.6cm, transform shape = true] {};
        }
    },
    pics/cameraTwo/.style n args={2}{
        code = { %
            \draw (0,0) node[draw,fill=#1, thick, minimum width=2cm,minimum height=1cm, transform shape = true] {};
            \draw (1.125,0.025) node[draw,fill=#2, thick, minimum width=0.2cm,minimum height=0.6cm, transform shape = true] {};
        }
    },
    pics/mybus/.style n args={3}{
        code = { %
        \draw[thick, draw=gray, fill=black,transform shape = true] (0,0) circle(20pt);
        \draw[thick,fill=gray, draw=white,transform shape = true] (0,0) circle(12pt);
        \draw[thick, fill=black,draw=white,transform shape = true] (0,0) circle(4pt);
        \draw[decorate,decoration={ticks}, fill=white,text=white,color=white,transform shape = true] (0,0) circle (8pt);
        \draw[thick, draw=gray, fill=black,transform shape = true] (4,0) circle(20pt);
        \draw[thick,fill=gray, draw=white,transform shape = true] (4,0) circle(12pt);
        \draw[thick, fill=black,draw=white,transform shape = true] (4,0) circle(4pt);
        \draw[decorate,decoration={ticks}, fill=white,text=white,color=white,transform shape = true] (4,0) circle (8pt);
        \draw[thick,draw=#3,fill=#1,transform shape = true] (-2,0) -- ++(1,0) -- ++(0.5,0.5) -- ++(1,0) -- ++(0.5,-0.5) --
        ++(2,0) -- ++(0.5,0.5) -- ++(1,0) -- ++(0.5,-0.5) -- ++(1,0) -- ++(0,3.5) -- ++(-1,0.5) -- ++(-6,0) -- ++(-1,-0.5) -- ++(0,-3.5) -- ++(0.5,0) ;
        \draw[thick,draw=#3,fill=#2,transform shape = true] (-2,1.5) -- ++(2,0) -- ++(0,2) -- ++(-2,0) -- ++(0,-2);
        \draw[thick,draw=#3,fill=#2,transform shape = true] (4,1.5) -- ++(2,0) -- ++(0,2) -- ++(-2,0) -- ++(0,-2);
        \draw[thick,draw=#3,fill=#2,transform shape = true] (0.5,1.5) -- ++(3,0) -- ++(0,2) -- ++(-3,0) -- ++(0,-2);

        }
    },
    pics/busstop/.style n args={1}{
        code = { %
            \draw [draw=black!60!green, line width=1mm,fill=yellow,transform shape = true] (0,0) circle (12pt);
            \draw (0,0) node[text=black!60!green,transform shape = true] {\Large \textbf{#1}};
        }
    },
    pics/caution/.style n args={1}{
        code = { %
            \draw (0,0) node[regular polygon, regular polygon sides=3, draw=red,line width=1mm, fill=white!20,
            text width=1em, text badly centered,
            inner sep=1pt, rounded corners,transform shape = true] {\Large{\textbf{#1}}};
        }
    },
    pics/trffclight/.style n args={3}{
        code = { %
            \draw (0,0) node[draw,fill=black, thick, minimum width=1cm,minimum height=2cm,transform shape = true] {};
            \draw [fill=#1,transform shape = true] (0,0.65) circle (8pt);
            \draw [fill=#2,transform shape = true] (0,0) circle (8pt);
            \draw [fill=#3,transform shape = true] (0,-0.65) circle (8pt);
        }
    },
    pics/server/.style n args={3}{
        code = { %
            \draw[thick,draw=black,fill=#1,transform shape = true] (0,0) -- (1,0) -- (1,2) -- (0,2) -- (0,0);
            \draw[thick,draw=black,fill=#2,transform shape = true] (0,2) --(0.5,2.5) -- (1.5,2.5) -- (1,2) -- (0,2);
            \draw[thick,draw=black,fill=#3,transform shape = true] (1,0) --(1,2) -- (1.5, 2.5) -- (1.5, 0.5) -- (1,0);
            \draw[thick, fill=white,draw=black,transform shape = true] (0.5,0.6) circle(3pt);
            \draw[draw=black,transform shape = true] (0.1,0.1) --(0.9,0.1);
            \draw[draw=black,transform shape = true] (0.1,0.3) --(0.9,0.3);
            \draw[draw=black,transform shape = true] (0.1,0.2) --(0.9,0.2);
            \draw[thick, fill=black,draw=gray!75!black,transform shape = true] (0.2,1.8) circle(1pt);
            \draw[thick, fill=black,draw=gray!75!black,transform shape = true] (0.4,1.8) circle(1pt);
            \draw[line width=2pt,transform shape = true] (0.75,1.8) -- (0.9,1.8);
            \draw[line width=2pt,transform shape = true] (0.55,1.8) -- (0.7,1.8);
            \draw[line width=2pt,transform shape = true] (0.1,1.6) -- ( 0.9,1.6);
        }
    },
    pics/cellularsite/.style n args={1}{
        code = { %
            \draw[line width=1pt,#1,transform shape = true] (0,0) -- ( 1,-4) -- (0,-3) -- (-1,-4) -- (0,0);
            \draw[line width=1pt,#1,transform shape = true] (-0.75,-3) -- (0.75,-3) -- (-0.5,-2) -- (0.5,-2) -- (-0.75,-3);
            \draw[line width=1pt,#1,transform shape = true] (-0.5,-2) -- ( 0.5,-2) -- (-0.25,-1) -- (0.25,-1) -- (-0.5,-2);
            \draw[line width=1pt,#1,transform shape = true] (-0.5,-2) -- ( -1.25,-2) -- (-0.375,-1.5);
            \draw[line width=1pt,#1,transform shape = true] (0.5,-2) -- ( 1.25,-2) -- (0.375,-1.5);
        }
    },
    pics/odometry/.style n args={3}{
        code = { %
            \draw[draw=black,transform shape = true] (0,0) circle(16pt);
            \draw[decorate,thick,decoration={ticks,segment length=4pt},color=black,transform shape = true] (0,0) circle (16pt);
            \draw[->, thick,transform shape = true] (-30:1cm) arc[radius=1, start angle=-30, end angle=60];
        }
    },
    pics/chipOne/.style n args={1}{
        code = { %
            \draw (0,0) node[draw,fill=black!60!green, thick, minimum width=2cm,minimum height=1cm, transform shape = true] {};
            \draw[decorate,decoration={border,amplitude=0.2cm,angle=90, segment length=4pt},transform shape = true, ultra thick] (-0.75,0) - - (1.25,0);
        }
    },
    pics/cloud/.style n args={1}{
        code = { %
            \draw node[cloud, cloud puffs=15.7, cloud ignores aspect, minimum width=2.5cm, minimum height=1.5cm, align=center, draw, transform shape = true] {#1};
        }
    },
    pics/vehicle/.style n args={2}{
        code = { %
            \draw[thick, draw=gray, fill=black,transform shape = true] (1,0) circle(20pt);
            \draw[thick,fill=gray, draw=white,transform shape = true] (1,0) circle(12pt);
            \draw[thick, fill=black,draw=white,transform shape = true] (1,0) circle(4pt);
            \draw[decorate,decoration={ticks}, fill=white,text=white,color=white,transform shape = true] (1,0) circle (8pt);
            \draw[thick, draw=gray, fill=black,transform shape = true] (5,0) circle(20pt);
            \draw[thick,fill=gray, draw=white,transform shape = true] (5,0) circle(12pt);
            \draw[thick, fill=black,draw=white,transform shape = true] (5,0) circle(4pt);
            \draw[decorate,decoration={ticks}, fill=white,text=white,color=white,transform shape = true] (5,0) circle (8pt);
            \draw[thick,draw=black,fill=#1,transform shape = true] (0,0) -- ++(0.5,0.5) -- ++(1,0) -- ++(0.5,-0.5) -- ++(2,0) -- ++(0.5,0.5) -- ++(1,0) -- ++(0.5,-0.5) -- ++(0.5,0) -- ++(0.3,0.5) -- ++(0,1) -- ++(-0.75,0.3) -- ++(-1.75,0.2) -- ++(-0.5,1) -- ++(-0.5,0.25) -- ++(-2.75,0) -- ++(-1.25,-0.35) -- ++(0,-2.15) -- ++(0.5,-0.75) -- ++(0.25,0);
            \draw[thick,draw=black,fill=#2,transform shape = true] (2.5,2) -- ++(1.5,0) -- ++(-0.5,1) -- ++(-1,0) -- ++(0,-1);
            \draw[thick,draw=black,fill=#2,transform shape = true] (1,2) -- ++(1.25,0) -- ++(0,1) -- ++(-1.25,0) -- ++(0,-1);
            \draw[thick,draw=black,fill=#2,transform shape = true] (-0.5,2) -- ++(1.25,0) -- ++(0,1) -- ++(-1.25,-0.25) -- ++(0,-0.75);
        }
    },
    pics/fuelcharge/.style n args={1}{
        code = {
            \draw (0,0) node[draw,fill=black, thick, minimum width=1cm,minimum height=1.8cm,transform shape = true] {};
            \draw (-0.025,0.5) node[draw,fill=white, thick, minimum width=0.6cm,minimum height=0.5cm,transform shape = true] {};
            \draw (0,-0.8) node[draw,fill=black, thick, minimum width=1.2cm,minimum height=0.25cm,transform shape = true] {};
            \draw[ultra thick] (0.5,-0.1) .. controls (1,-0.3) and (1,0.5) .. (0.65,0.75);
            \draw [draw=black, line width=0.5mm,transform shape = true] (0.75,0.5) circle (2pt);
        }
    },
    pics/switch/.style n args={1}{
        code = { %
            \draw[thick,draw=black,fill=#1,transform shape = true] (0,0) -- (2,0) -- (2,0.25) -- (0,0.25) -- (0,0);
            \draw[thick,draw=black,fill=#1,transform shape = true] (0,0.25) --(2,0.25) -- (2.3,0.4) -- (0.3,0.4) -- (0,0.25);
            \draw[thick,draw=black,fill=#1,transform shape = true] (2,0) -- (2,0.25) -- (2.3,0.4) -- (2.3, 0.15) -- (2,0);
            
            \draw[thick,draw=black,fill=black,transform shape = true] (0,0) -- ++(0.2,0) -- ++(0,0.1) -- ++(0.1,0) -- ++(0,-0.1) -- ++(0.2,0) -- ++(0,0.1) -- ++(0.1,0) -- ++(0,-0.1) -- ++(0.2,0) -- ++(0,0.1) -- ++(0.1,0) -- ++(0,-0.1) -- ++(0.2,0) -- ++(0,0.1) -- ++(0.1,0) -- ++(0,-0.1) -- ++(0.2,0) -- ++(0,0.1) -- ++(0.1,0) -- ++(0,-0.1) -- ++(0.2,0) -- ++(0,0.1) -- ++(0.1,0) -- ++(0,-0.1) -- ++(0.2,0) -- ++(0,0);
        }
    },
    pics/street/.style n args={1}{
        code = { %
            \draw[ultra thick,line width=5pt, transform shape = true] (0,1) -- (4,1) -- (4,2) -- (6,2) -- (6,-1) -- (-4,-1) -- (-4,1) -- (0,1);
            \draw[white,line width=0.5mm, loosely dashed,transform shape = true] (0,1) -- (4,1) -- (4,2) -- (6,2) -- (6,-1) -- (-4,-1) -- (-4,1) -- (0,1);
        }
    },
    pics/satelite/.style n args={1}{
        code = { %
            \draw[thick,draw=black,transform shape = true] (0,0) -- ++(2,0) -- ++(0,1) -- ++(-2,0) -- ++(0,-1);
            \draw[thick,draw=black,fill=black,transform shape = true] (1,1) -- ++(0,0.5);
            \draw[thick, draw=black,fill=#1,transform shape = true] (0,1.5) -- ++(2,0) -- ++(0,4) -- ++(-2,0) -- ++(0,-4);
            \draw[thick, draw=black,transform shape = true] (0,1.5) -- ++(2,0) -- ++(0,0.5) -- ++(-2,0) -- ++(0,0.5) -- ++(2,0) -- ++(0,0.5) -- ++(-2,0) -- ++(0,0.5) -- ++(2,0) -- ++(0,0.5) -- ++(-2,0) -- ++(0,0.5) -- ++(2,0) -- ++(0,0.5) -- ++(-2,0) -- ++(0,0.5) -- ++ (2,0) -- ++ (0,-4) -- ++(-2,0) -- ++ (0,4) -- ++ (1,0) -- ++ (0,-4);
            \draw[thick,draw=black,fill=black,transform shape = true] (1,0) -- ++(0,-0.5);
            \draw[thick, draw=black, fill=#1,transform shape = true] (0,-0.5) -- ++(2,0) -- ++(0,-4) -- ++(-2,0) -- ++(0,4);
            \draw[thick, draw=black,transform shape = true] (0,-0.5) -- ++(2,0) -- ++(0,-0.5) -- ++(-2,0) -- ++(0,-0.5) -- ++(2,0) -- ++(0,-0.5) -- ++(-2,0) -- ++(0,-0.5) -- ++(2,0) -- ++(0,-0.5) -- ++(-2,0) -- ++(0,-0.5) -- ++(2,0) -- ++(0,-0.5) -- ++(-2,0) -- ++(0,-0.5) -- ++ (2,0) -- ++ (0,4) -- ++(-2,0) -- ++ (0,-4) -- ++ (1,0) -- ++ (0,4);
            \draw[thick,draw=black,fill=black,transform shape = true] (-0.5,0.25) -- ++(0.5,0) -- ++ (0,0.5) -- ++(-0.5,0) -- ++(0,-0.5);
            \draw (2.5,1) arc (90:270:0.5cm) -- ++(0,1);
            \path (2.75,0.5) edge [decoration={expanding waves,angle=60,segment length=3mm},decorate,draw,color=green] (4,0.5);
            \draw[thick,draw=black,transform shape = true] (2.5,0.5) -- ++(0.25,0);
        }
    },
    pics/textrect/.style n args={2}{
        code = { %
            \draw[thick, fill=#2, inner sep= 5pt,transform shape = true] (0,0) rectangle node[pos=.5]{\textbf{#1}} (4,1);
        }
    },
    pics/box/.style n args={2}{
        code = { %
            \draw[thick, draw=black,transform shape = true] (0,0) rectangle (#1,#2);
        }
    },
    pics/smartphone/.style n args={2}{
        code = { %
            \draw[thick, draw=black, fill=gray, transform shape = true] (0,0) rectangle (2,4);
            \draw[thick, draw=black, fill=white, transform shape = true] (0.1,0.6) rectangle (1.9,3.9);
            \draw[thick, draw=black, fill=white, transform shape = true] (1,0.3) circle (0.2cm);
        }
    },
    pics/basestation/.style n args={2}{
        code = { %
            \draw (0,0.75) arc (210:390:0.5cm) -- ++(0,0) -- ++(-0.9,-0.5);
            \draw[thick, fill,draw=black,transform shape = true] (0.3,0) -- ++(0.7,0) -- ++(-0.35,0.55) -- ++(-0.35,-0.55);
            \draw[thick,draw=black,transform shape = true] (0.425,1) -- ++(-0.15,0.25);
            \draw[thick, fill=black] (0.3,1.2) circle(2pt);
        }
    }
}
\usepackage{lipsum}
\usepackage{textcomp}
\usepackage{pgfplots}
\usepackage{xintexpr}
\usepackage{scalerel}
\usepackage{tikz}
\usetikzlibrary{svg.path}

\usepackage{array,booktabs,ragged2e}
\newcolumntype{R}[1]{>{\RaggedLeft\arraybackslash}p{#1}}
\newcolumntype{L}[1]{>{\RaggedRight\arraybackslash}p{#1}}

\usepackage{filecontents}

\usepackage{amssymb}
\usepackage{amsmath}
\usepackage{changes}
\definechangesauthor[name=Mudassar, color=red]{MA}
\definechangesauthor[name=Sebastian, color=blue]{SS}

\usepackage{ulem} 

\newcommand\orcidicon[1]{\href{https://orcid.org/#1}{\mbox{\scalerel*{
\begin{tikzpicture}[yscale=-1,transform shape]
\pic{orcidlogo};
\end{tikzpicture}
}{|}}}}

\usepackage{hyperref} 

\begin{document}

%

\title{Quantitative System-Level Security Verification of the IoV Infrastructure}
%
%
%

\author{Jan Lauinger \orcidicon{0000-0002-4917-1850}\,,~\IEEEmembership{IEEE, Member},
        Mudassar Aslam \orcidicon{0000-0003-3223-4234}\,,
        Mohammad Hamad \orcidicon{0000-0002-9049-7254}\,,
        Shahid Raza \orcidicon{0000-0001-8192-0893}\,,~\IEEEmembership{IEEE, Senior Member},
        and Sebastian Steinhorst \orcidicon{0000-0002-4096-2584}\,,~\IEEEmembership{IEEE, Senior Member}
\thanks{J. Lauinger, M. Hamad, S. Steinhorst are with the Department of Electrical and Computer Engineering, Technical University of Munich, Munich, 80333 Germany, E-mail: (jan.lauinger@tum.de, mohammad.hamad@tum.de, sebastian.steinhorst@tum.de).}
\thanks{M. Aslam and S. Raza are with the Cybersecurity Unit, RISE Research Institutes of Sweden, Stockholm, 16440 Sweden, E-mail: (mudassar.aslam@ri.se, shahid.raza@ri.se).}
\thanks{This work has received funding by the European Union’s Horizon 2020 Research and Innovation Programme through the nIoVe project (https://www.niove.eu/) under grant agreement no. 833742, and the CONCORDIA project (https://concordia-h2020.eu/) under the grant agreement no. 830927.}
\thanks{With the support of the Technische Universität München - Institute for Advanced Study, funded by the German Excellence Initiative and the European Union Seventh Framework Programme under grant agreement no. 291763.}
\thanks{Manuscript received July 14, 2020; }} 

%
%

\markboth{IEEE Journal}%
{Shell \MakeLowercase{\textit{et al.}}: Bare Demo of IEEEtran.cls for IEEE Journals}
%








\maketitle


\begin{abstract}
The Internet of Vehicles (IoV) equips vehicles with connectivity to the Internet and the Internet of Things (IoT) to support modern applications such as autonomous driving. However, the consolidation of complex computing domains of vehicles, the Internet, and the IoT limits the applicability of tailored security solutions. In this paper, we propose a new methodology to quantitatively verify the security of single or system-level assets of the IoV infrastructure. In detail, our methodology decomposes assets of the IoV infrastructure with the help of reference sub-architectures and the 4+1 view model analysis to map identified assets into data, software, networking, and hardware categories. This analysis includes a custom threat modeling concept to perform parameterization of Common Vulnerability Scoring System (CVSS) scores per view model domain. As a result, our methodology is able to allocate assets from attack paths to view model domains. This equips assets of attack paths with our IoV-driven CVSS scores. Our CVSS scores assess the attack likelihood which we use for Markov Chain transition probabilities. This way, we quantitatively verify system-level security among a set of IoV assets.
Our results show that our methodology applies to arbitrary IoV attack paths. Based on our parameterization of CVSS scores and our selection of use cases, remote attacks are less likely to compromise location data compared to attacks from close proximity for authorized and unauthorized attackers respectively.

\end{abstract}


%
\IEEEpeerreviewmaketitle

\section{Introduction}


\IEEEPARstart{N}{ew} connectivity capabilities in the \ac{iov} provide vehicles with access to the infrastructure of the Internet.
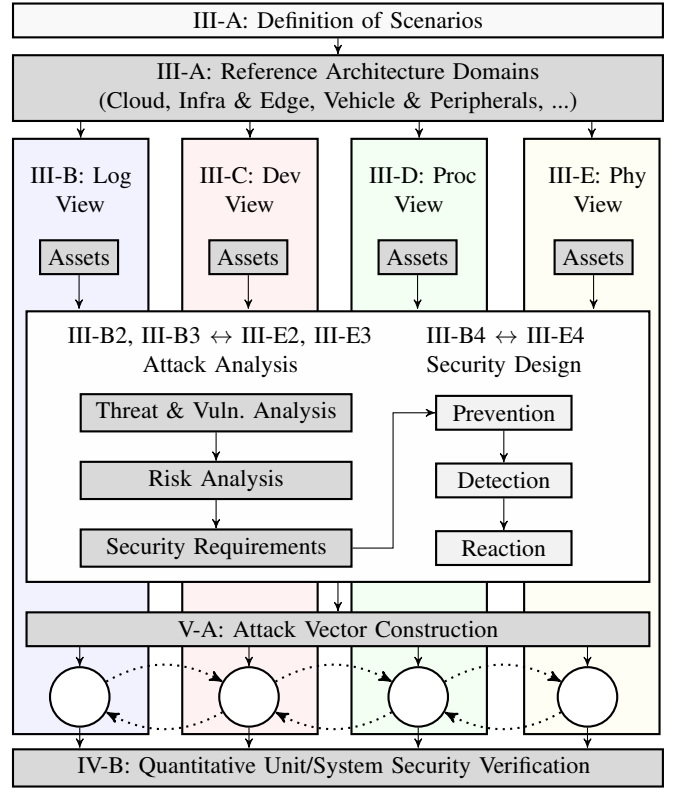
\begin{figure}[!ht]
    \centering
    \begin{tikzpicture}[scale=0.9, ->, >=stealth', node distance=2.5cm]
    
    \node (sec1) at (-1,10) [rectangle, anchor=west, thick, draw=black, fill=white!95!gray, minimum width=9.6cm, minimum height=0.5cm, transform shape = true, align=left] {\ref{scenariosec}: Definition of Scenarios};
    \node (sec2) at (-1,9) [rectangle, anchor=west, thick, draw=black, fill=white!70!gray, minimum width=9.6cm, minimum height=0.5cm, transform shape = true, align=left] {\ \ \ \ \ \ \ \ref{scenariosec}: Reference Architecture Domains\\ (Cloud, Infra \& Edge, Vehicle \& Peripherals, ...) };

    \node (abox1) at (-1,3.85) [draw, anchor=west, thick, fill=white!95!blue, align=left, minimum height=8.8cm, minimum width=2cm, transform shape = true] {};
    \node (as1) at (-0.85,7.5) [transform shape = true, align=left, anchor=west] {\ref{section3lv}: Log \\ \ \ \ View};
    \node (as1) at (-0.6,6.5) [rectangle, anchor=west, thick, draw=black, fill=white!70!gray, minimum width=1cm, minimum height=0.5cm, transform shape = true, align=left] {Assets};
    
    \node (abox2) at (1.5,3.85) [draw, anchor=west, thick, fill=white!95!red, align=left, minimum height=8.8cm, minimum width=2cm, transform shape = true] {};
    \node (vuln) at (1.65,7.5) [transform shape = true, align=left, anchor=west] {\ref{section3dv}: Dev \\ \ \ \ View};
    \node (as2) at (1.9,6.5) [rectangle, anchor=west, thick, draw=black, fill=white!70!gray, minimum width=1cm, minimum height=0.5cm, transform shape = true, align=left] {Assets};
    
    \node (abox3) at (4,3.85) [draw, anchor=west, thick, fill=white!95!green, align=left, minimum height=8.8cm, minimum width=2cm, transform shape = true] {};
    \node (vuln) at (4.15,7.5) [transform shape = true, align=left, anchor=west] {\ref{section3prv}: Proc \\ \ \ \ View};
    \node (as3) at (4.4,6.5) [rectangle, anchor=west, thick, draw=black, fill=white!70!gray, minimum width=1cm, minimum height=0.5cm, transform shape = true, align=left] {Assets};
    
    \node (abox4) at (6.55,3.85) [draw, anchor=west, thick, fill=white!95!yellow, align=left, minimum height=8.8cm, minimum width=2cm, transform shape = true] {};
    \node (vuln) at (6.8,7.5) [transform shape = true, align=left, anchor=west] {\ref{section3phv}: Phy \\ \ \ \ View};
    \node (as4) at (7,6.5) [rectangle, anchor=west, thick, draw=black, fill=white!70!gray, minimum width=1cm, minimum height=0.5cm, transform shape = true, align=left] {Assets};

    \node (abox5) at (-0.8,3.7) [draw, anchor=west, thick, fill=white!99!gray, align=left, minimum height=4cm, minimum width=9.2cm, transform shape = true] {};
    \node (vuln) at (-0.3,5.1) [transform shape = true, align=left, anchor=west] {\ref{datathreatvulnerability}, \ref{section3rasr} $\leftrightarrow$ \ref{section3phytva}, \ref{section3phyrasr} \\ \ \ \ \ \ \ \ \ \ Attack Analysis};
    \node (vuln) at (5,5.1) [transform shape = true, align=left, anchor=west] {\ref{section3sc} $\leftrightarrow$ \ref{section3physc} \\Security Design};
    
    \node (at1) at (0,4.2) [rectangle, anchor=west, thick, draw=black, fill=white!70!gray, minimum width=4cm, minimum height=0.5cm, transform shape = true, align=left] {Threat \& Vuln. Analysis};
    \node (at2) at (0,3.2) [rectangle, anchor=west, thick, draw=black, fill=white!70!gray, minimum width=4cm, minimum height=0.5cm, transform shape = true, align=left] {Risk Analysis};
    \node (at3) at (0,2.2) [rectangle, anchor=west, thick, draw=black, fill=white!70!gray, minimum width=4cm, minimum height=0.5cm, transform shape = true, align=left] {Security Requirements};

    \node (secd1) at (5.25,4.2) [rectangle, anchor=west, thick, draw=black, fill=white!90!gray, minimum width=2cm, minimum height=0.5cm, transform shape = true, align=left] {Prevention};
    \node (secd2) at (5.25,3.2) [rectangle, anchor=west, thick, draw=black, fill=white!90!gray, minimum width=2cm, minimum height=0.5cm, transform shape = true, align=left] {Detection};
    \node (secd3) at (5.25,2.2) [rectangle, anchor=west, thick, draw=black, fill=white!90!gray, minimum width=2cm, minimum height=0.5cm, transform shape = true, align=left] {Reaction};

    \node (avc1) at (-0.8,1) [rectangle, anchor=west, thick, draw=black, fill=white!70!gray, minimum width=9.2cm, minimum height=0.5cm, transform shape = true, align=left] {\ref{section4avc}: Attack Vector Construction};
    
    \tikzstyle{every state}=[fill=white,draw=black,thick,text=black,scale=1,transform shape = true]
            \node[state]    (A)               {};
            \node[state]    (B)[right of=A]   {};
            \node[state]    (C)[right of=B]   {};
            \node[state]    (D)[right of=C]   {};
            \path
            (A) edge[bend left, dotted, thick]     node{}     (B)
            (B) edge[bend left, dotted, thick]    node{}           (C)
                edge[bend left, below, dotted, thick]    node{}           (A)
            (C) edge[bend left, dotted, thick]    node{}   (D)
                edge[bend left, below, dotted, thick]     node{}   (B)
            (D) edge[bend left, dotted, thick, below]    node{}    (C);

    \node (el1) at (-1,-1.05) [rectangle, anchor=west, thick, draw=black, fill=white!70!gray, minimum width=9.6cm, minimum height=0.5cm, transform shape = true, align=left] {\ref{evalmodel}: Quantitative Unit/System Security Verification };
    
    \draw [->] (sec1.south) -- (sec2.north);
    
    \draw [->] (as1.south) -- ++(0,-0.5);
    \draw [->] (as2.south) -- ++(0,-0.5);
    \draw [->] (as3.south) -- ++(0,-0.5);
    \draw [->] (as4.south) -- ++(0,-0.5);

    \draw [<-] (abox1.north) -- ++(0,0.25);
    \draw [<-] (abox2.north) -- ++(0,0.25);
    \draw [<-] (abox3.north) -- ++(0,0.25);
    \draw [<-] (abox4.north) -- ++(0,0.25);
    
    \draw [->] (at1.south) -- (at2.north);
    \draw [->] (at2.south) -- (at3.north);
    
    \draw [->] (at3.east) -- ++(0.6,0) -- ++(0,2) -- (secd1.west);
    
    \draw [->] (secd1.south) -- (secd2.north);
    \draw [->] (secd2.south) -- (secd3.north);
    
    \draw [->] (abox5.south) -- (avc1.north);
    
    \draw [<-] (A.north) -- ++(0,0.3);
    \draw [<-] (B.north) -- ++(0,0.3);
    \draw [<-] (C.north) -- ++(0,0.3);
    \draw [<-] (D.north) -- ++(0,0.3);
    
    \draw [->] (A.south) -- ++(0,-0.3);
    \draw [->] (B.south) -- ++(0,-0.3);
    \draw [->] (C.south) -- ++(0,-0.3);
    \draw [->] (D.south) -- ++(0,-0.3);

    \end{tikzpicture}
  \caption{High-level illustration, including paper section references, of the proposed methodology for quantitative system-level security verification.}
\label{fig:mainsecuritydesigntailored}
\end{figure}
As a result, upcoming services around connected vehicles access new forms of data for enhanced driving experiences, safety, and automation such as autonomous decision making over maneuvers \cite{kaiwartya2016internet}.
Simultaneously, increasing connectivity causes an increase in complexity which, from a security perspective, opens up a larger attack surface. 
Attackers, who successfully compromise vulnerabilities of the \ac{iov} infrastructure, face new opportunities to remotely interfere with vehicles. 
As a direct consequence, the potential of attacks that affect vehicle safety by accident or on purpose increases \cite{wong2017uber}--\cite{eiza2017driving}. 

For the reason that jeopardized safety-critical systems threaten \ac{iov} acceptance, the investigation of holistic \ac{iov} security concepts represents a common interest of \ac{iov} stakeholders \cite{rizvi2017threat}.
Despite the existence of new and comprehensive security solutions for the \ac{iov}, they remain in an early development stage \cite{schmittner2018status}, or face difficulties with administrative, legal, or technical development \cite{parkinson2017cyber}.
Thus, the interplay of different technological domains in the \ac{iov} demand tailored, automated, dynamic, and adaptive security solutions.

To address this challenge and to evaluate new security concepts for assets of the \ac{iov} infrastructure, we propose a new methodology that allows to quantitatively verify system-level security solutions.
Our methodology requires the definition of attack paths to define assets for the security verification. Additionally, our methodology requires an analysis of the IoV reference architecture to allocate, equip, and assess identified assets.

In order to analyze complex assets in a structured way, reference models, layers, or view models provide ways to categorize the structure of an asset by highlighting different groups of aspects. 
The 4+1 architectural view model, used in our work, provides the logical, process, developer, and physical views to analyze data, communication, libraries and dependencies, and hardware aspects respectively \cite{kruchten19954+}.
We leverage the separated analysis of the \ac{iov} assets per view to (1) identify assets of the \ac{iov} infrastructure and (2) to accurately map attacks as well as defense mechanisms to assets. As a result, we can label properties of \ac{cvss} scores for \ac{iov} assets, respecting each view category individually. This view model-based attack analysis allows us to identify security measures per asset that an attacker needs to compromise.

An attack is successful if the attacker exploits vulnerabilities or if the attacker breaches security mechanisms \cite{elahi2011security}. To reach the goal of an attack path, an attacker is required to perform successful attacks repetitively. In order to model the attacker perspective at different stages as well as quantitatively verify system security, our work leverages state transitions probabilities of Markov Chains. In this context, state transitions represent attacker stages of attack trees. To assess each individual stage of an attack path, we leverage the vulnerability, risk, and security analysis based on \ac{cvss} scores. 
The structure of Markov Models enables our quantitative security verification of \ac{iov} assets as well as opportunities to verify system-level security of multiple assets that are part of attack path \cite{cheng2012research}.

To recap the consecutive steps of our methodology, Figure \ref{fig:mainsecuritydesigntailored} indicates each step that are necessary for the quantitative system-level security verification of the \ac{iov} infrastructure.
At the same time, Figure \ref{fig:mainsecuritydesigntailored} refers to the sections of our work which apply the respective analysis.
With a general focus on the \ac{iov} location service application (see Section \ref{scenariosec}), we leverage sub-architectures, defined in the work \cite{maple2019connected}, to model \ac{iov} system assets.
This measure reduces the complexity and facilitates our security analysis.
Section \ref{section3} applies the 4+1 view model analysis of the \ac{iov} architecture from a security perspective. Based on our knowledge, our work applies the 4+1 view model in the \ac{iov} security context for the first time.
Section \ref{threatmodeling} applies our agile threat modeling concept to handle dynamics of the \ac{iov} architecture during the assessment \cite{oyler2016security}--\cite{cruzes2018challenges}.
Section \ref{section3lv} to \ref{section3phv} investigate identified assets to determine \ac{iov}-specific \ac{cvss} vulnerability scores.
After selecting IoV attack paths in Section \ref{section4avc}, we take the \ac{iov}-driven \ac{cvss} scores as parameters to our Markov Chain model in Section \ref{section5}.

Our 4+1 view model analysis of the \ac{iov} infrastructure reveals attacks and security requirements per asset. The collected assumptions about existing attacks and security mechanisms enable us to define \ac{iov} driven \ac{cvss} scores. Based on our \ac{cvss} scores, it is possible to quantitatively verify system-level security of components that are part of different attack trees. Attack trees depend on our selection of existing \ac{iov} attacks that target location services. Our results ascertain less chances for remote attacks to compromise location data of \ac{cavs} compared to close proximity attacks. Apart from system-level security verification, our results show that it is possible to apply our methodology to multiple existing \ac{iov} attacks to achieve a comparable security assessment among components of the \ac{iov} infrastructure.

To sum up in bullet points, we contribute with: 
\begin{itemize}
    \item Applying the 4+1 view model analysis in the \ac{iov} security context for the first time to the best of our knowledge.
    \item Performing security risk assessment based on \ac{iov}-driven \ac{cvss} scores.
    \item Proposing an agile, modular, view model-based methodology to design and verify security concepts for \ac{iov} systems.
    \item Applying and evaluating our proposed methodology using existing \ac{iov} attacks targeting location services of \ac{cavs}.
\end{itemize}

\section{Background \& Related Work}
\label{section2}

\subsection{View Model Frameworks in the Security Context}
\label{viewmodels}

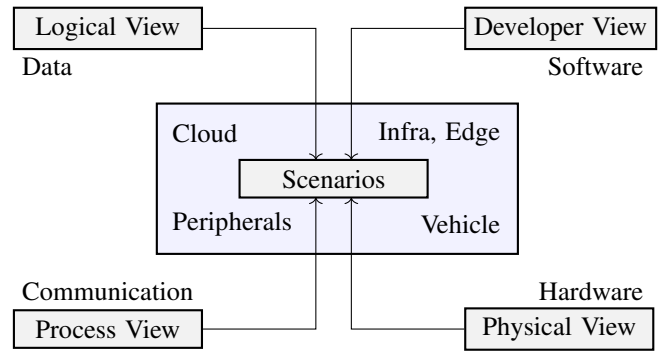
\begin{figure}[t]
\centering
\begin{tikzpicture}[scale=1]

\node (v1) at (-3,2) [scale=1, rectangle, anchor=west, thick, draw=black, fill=white!90!gray, minimum width=2.5cm, minimum height=0.5cm, transform shape = true, align=left] {Logical View};
\node (v12) at (-3,1.5) [transform shape = true, align=left, anchor=west] {Data};

\node (v2) at (3,2) [scale=1, rectangle, anchor=west, thick, draw=black, fill=white!90!gray, minimum width=2.5cm, minimum height=0.5cm, transform shape = true, align=left] {Developer View};
\node (v21) at (5.5,1.5) [transform shape = true, align=left, anchor=east] {Software};

\node (v3) at (-3,-2) [scale=1, rectangle, anchor=west, thick, draw=black, fill=white!90!gray, minimum width=2.5cm, minimum height=0.5cm, transform shape = true, align=left] {Process View};
\node (v31) at (-3,-1.5) [transform shape = true, align=left, anchor=west] {Communication};

\node (v4) at (3,-2) [scale=1, rectangle, anchor=west, thick, draw=black, fill=white!90!gray, minimum width=2.5cm, minimum height=0.5cm, transform shape = true, align=left] {Physical View};
\node (v41) at (5.5,-1.5) [transform shape = true, align=left, anchor=east] {Hardware};

\node (v5) at (-1.1,0) [draw, anchor=west, thick, fill=white!95!blue, align=left, minimum height=2cm, minimum width=4.8cm, transform shape = true] {};
\node (v6) at (0,0) [scale=1, rectangle, anchor=west, thick, draw=black, fill=white!90!gray, minimum width=2.5cm, minimum height=0.5cm, transform shape = true, align=left] {Scenarios};
\node (v7) at (3.6,0.6) [transform shape = true, align=left, anchor=east] {Infra, Edge};
\node (v8) at (-1,0.6) [transform shape = true, align=left, anchor=west] {Cloud};
\node (v9) at (3.6,-0.6) [transform shape = true, align=left, anchor=east] {Vehicle};
\node (v10) at (-1,-0.6) [transform shape = true, align=left, anchor=west] {Peripherals};

\draw [->] (v1.east) -- ++(1.5,0) -- ++(0,-1.75);
\draw [->] (v2.west) -- ++(-1.5,0) -- ++(0,-1.75);
\draw [->] (v3.east) -- ++(1.5,0) -- ++(0,1.75);
\draw [->] (v4.west) -- ++(-1.5,0) -- ++(0,1.75);


\end{tikzpicture}
  \caption{The 4+1 View Model in the \ac{iov} context.}
\label{fig:41IoVscenario}
\end{figure}

There are multiple view angles to analyze an \ac{iov} infrastructure \cite{karkhanis2018defining}. 
Considering functional, communication, implementation, enterprise, usage, information, physical, stakeholder, and user viewpoints all together is not beneficial regarding security analysis \cite{maple2019connected}. 
Categories may introduce either too much complexity and inconsistencies or, in essence, do not contribute to security-related purposes such as attack analysis. 
Hence, a sufficiently balanced portfolio of viewpoints increases the applicability of appropriate security concepts \cite{hughes2013quantitative}.
It is possible to balance the tradeoff between the applicability of security concepts and the complexity of reference architectures by utilizing the approach of the 4+1 view model which describes the architecture of a scenario using multiple abstract views \cite{kruchten19954+}.
Figure \ref{fig:41IoVscenario} shows the 4+1 view model in the \ac{iov} context together with common characteristics that apply in each of the views.
The abstraction levels of the view model enable the identification of security-relevant system boundaries and information flows.

Focusing on each view individually, the \emph{logical view} decomposes the system by leveraging principles of abstraction, encapsulation, and inheritance to describe end-user functionality and services. 
The \emph{process view} utilizes requirements such as performance, availability, concurrency, distribution, integrity, fault tolerance to map logical view abstractions into a process model of communicating processes and tasks which reveal computing load.

The \emph{development view} organizes software modules and their import and export relationships by considering rules such as partitioning, grouping, scope, decoupling, reuse, portability, communication, time dependence and reveals allocations of requirements, costs, and planning.

The \emph{physical view} model determines the physical components of computer networks, processors and interfaces.
Thereby, the physical view model considers non-functional system requirements such as performance, scalability, reliability, and availability to drive configuration, deployment and testing decisions of various nodes.
In essence, these properties determine capacity requirements of the physical architecture of the hardware.

Last, the scenario defines application procedures, sequences, and interactions, identifies validation, verification and illustration concepts, and marks the input to all view models.
In the context of threat modeling, the scenario definition is essential, as it reduces the complexity of the attack surface by prioritizing assets \cite{jaatun2019threat}.
As such, the scenarios enable a target-oriented modeling of the system and assets which represents the initial step of threat modeling and risk assessment.
As of today, standardized tool sets and development frameworks facilitate implementations of each view model individually.

\subsection{Threat Modeling}
\label{threatmodeling}

To identify the main characteristics among different threat modeling approaches, chapter two of the comprehensive work of Shostack \cite{shostack2014threat} introduces asset-centric, attack-centric, and software-centric strategies of threat modeling.
By iterating over the threat modeling methodologies of the survey of Hussain et al. \cite{hussain2014threat}, the \ac{stride} threat model, which identifies spoofing, tampering, repudiation, information disclosure, denial of service, and elevation of privilege as the main threats, counts as a software-centric threat model. 
Likewise, the \ac{stride} Average Model, Fuzzy Logic model and the Abuser Stories methodology \cite{shostack2014threat} utilize \ac{stride}.

Graphical threat modeling concepts, such as attack trees, model system assets or attacks at different attack propagation stages depending on the assignment of the security expert.
Hence, it is not possible to allocate attack trees to either of the three threat modeling approaches. 
Nevertheless, graphical concepts provide flexibility and extendibility and fit into iterative threat modeling procedures.

Attack libraries, such as the \ac{capec} \cite{barnum2008common} or the Intel \ac{tal} \cite{casey2007threat}, address the attributes of the attacker and represent attacker-centric models. 
The automotive-compatible \ac{tara} model marks another attacker-centric model that is based on different threat-related libraries \cite{karahasanovic2017adapting}.
To close the scope of approaches, the asset-centric \ac{octave} \cite{albert2001octave} and \ac{tvra} \cite{etsi2010intelligent} models analyze threats, risks, and vulnerabilities together.

\begin{figure}[t]
    \centering
    \begin{tikzpicture}[scale=0.9]
    
    

    \node (el1) at (0,0) [rectangle, anchor=west, thick, draw=black, fill=white!70!gray, minimum width=2.75cm, minimum height=1cm, transform shape = true, align=center] {System \\Modeling};
    \node (el2) at (3,0) [rectangle, anchor=west, thick, draw=black, fill=white!70!gray, minimum width=2.5cm, minimum height=1cm, transform shape = true, align=center] {Threat Analysis};
    \node (el3) at (5.75,0) [rectangle, anchor=west, thick, draw=black, fill=white!70!gray, minimum width=3cm, minimum height=1cm, transform shape = true, align=center] {Vulnerability \\Analysis};
    \node (el4) at (5.75,-1.25) [rectangle, anchor=west, thick, draw=black, fill=white!70!gray, minimum width=3cm, minimum height=1cm, transform shape = true, align=center] {Risk Analysis};
    \node (el5) at (3,-1.25) [rectangle, anchor=west, thick, draw=black, fill=white!70!gray, minimum width=2.5cm, minimum height=1cm, transform shape = true, align=center] {Security \\Requirements};
    \node (el6) at (0,-1.25) [rectangle, anchor=west, thick, draw=black, fill=white!70!gray, minimum width=2.75cm, minimum height=1cm, transform shape = true, align=center] {Tailored Security \\Design};
    
    \draw [->] (el1) -- (el2);
    \draw [->] (el2) -- (el3);
    \draw [->] (el3.east) -- ++(0.25,0) -- ++(0, -1.25) -- (el4.east);
    \draw [->] (el4) -- (el5);
    \draw [->] (el5) -- (el6);
    \draw [->] (el6.west) -- ++ (-0.25,0) -- ++(0,1.25) -- (el1.west);

    \end{tikzpicture}
  \caption{Agile threat modeling approach for the \ac{iov}.} 
\label{fig:iovthreatmodel}
\end{figure}
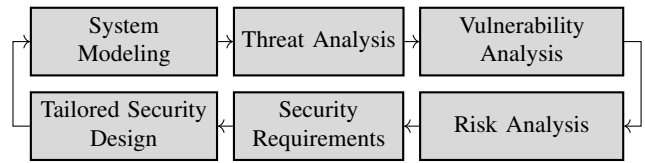

The work of \cite{karahasanovic2017adapting} and \cite{macher2016review} address the versatility and applicability of threat modeling approaches where \cite{karahasanovic2017adapting} proposes a tailored procedure of threat modeling for the \ac{iov}. 
This procedure adapts the \ac{tara} and \ac{stride} strategy.
Based on the work in \cite{karahasanovic2017adapting} and due to overlapping threat modeling strategies, the threat modeling of this work follows the strategy of Figure \ref{fig:iovthreatmodel}.
The strategy of the threat model of Figure \ref{fig:iovthreatmodel} represents an iterative process of system modeling, threat, vulnerability, and risk analysis, security requirement definition, and tailored security design.
This concept aligns with the proposed threat modeling procedures of \cite{shostack2014threat}, \cite{karahasanovic2017adapting}, and \cite{islam2016risk}.

To clarify the statement, \cite{shostack2014threat} relies on the four step framework of system modeling, threat analysis, threat mitigation, and validation.
In \cite{karahasanovic2017adapting}, the adapted \ac{tara} model lists threat-agent risk analysis, high-level threat-agent risk evaluation, attack method and exposure analysis, and strategy design to target exposures. 
The adapted \ac{tal}, \ac{mol}, and \ac{cel} drive this approach.

The work of \cite{islam2016risk} introduces a risk assessment framework for the automotive domain. 
The strategy relates to threat modeling approaches and consists of system definition, threat analysis, risk assessment, and definition of security requirements.
Their threat analysis identifies assets before the actual threats.
Moreover, the risk assessment block comprises threat and impact level estimation and security level determination.
Finally, the work of Hamad et al. \cite{hamad2020savta} combines all aspects of threat modeling, attack tree construction, and risk assessment in a comprehensive threat modeling approach tailored to vehicles. 

In contrast to general threat modeling approaches, such as fuzzy, probabilistic, tree-based, graphical, and legacy threat modeling, our work follows the agile threat modeling approach. 
The iterative nature of agile threat modeling provides the necessary flexibility of security analysis for the constantly evolving \ac{iov} domain.
With this approach, updates of security goals and requirements remain customizable \cite{cruzes2018challenges}. 
Software security analysts have the possibility to iteratively decrease abstraction levels, or change the quantification of scenarios.

Moreover, our approach of threat modeling of Figure \ref{fig:iovthreatmodel} includes a final block of tailored security design. 
The reason for this is the work of Xiong et al. \cite{xiong2019threat}, which states the design of a tailored security concept as future work, and the requirement definition of a mitigation concept in \cite{pacheco2016iot}.

Furthermore, the structures of the systematic threat modeling approach of \cite{ma2016threat} derive from the prominent \ac{iso26262} \cite{iso201126262} and \ac{saej3061} \cite{vcse2016j3061} standards which define a combination of \ac{tara} and \ac{stride} for risk assessment.
Our work relates in the way that it analyzes a high-level as well as in-depth details of modules and the implementation of the \ac{iov} architecture through the 4+1 view model analysis.
Likewise, our work identifies the security requirements of assets and provides a methodology to validate and verify security requirement effectiveness.



\subsection{System Design for Security Verification}

The work of Xiong et al. \cite{xiong2019threat} enhances threat modeling with probabilistic attack simulations that are based on networking graphs with attack paths. 
Their work builds upon the in-vehicle 2014 Jeep Cherokee \ac{can} network model of \cite{miller2014survey} and utilizes the software tool securiCAD for automated attack modeling and risk assessment.
The attack simulations based on the attack path incorporate attack types, vulnerabilities, and countermeasures at every propagation stage and manage to evaluate \ac{ttc} behavior.
The findings of this work demand more tailored definitions of meta-models, reference architectures, investigation of countermeasures, security architectures, validation of the approach through case studies, and quantitative input to the quantitative and probabilistic security metric of \ac{ttc}.

The work of Iqbal et al. \cite{iqbal2018context} describes the transition of the traditional \ac{iov} architecture into a data-driven-intelligent framework.
Their framework translates the architecture into data collection, preprocessing, data analysis, service, and application layers. 
Security, privacy, and trust management affect all layers.
Last, the approach of Zou et al. \cite{zou2017research} proposes an architecture that keeps a security monitoring system, threat intelligence, and networking security modules at the bottom layer. 
Validation and verification services build on top of the lowest layer. Defense, reinforcement, and response systems complete their so-called 360 connected vehicle safety framework.

To address the outcomes of \cite{xiong2019threat}, our work reproduces their threat modeling concept with the following changes.
Regarding the reference architecture, our work leverages the outcomes of the \ac{iov} reference model architecture analysis of \cite{maple2019connected}. 
Based on this model and using a scenario, we apply the 4+1 view model to break down assets to extract detailed vulnerability properties. Our abstraction concept differentiates between hardware, software, networking, and data and enables mappings of attacks and defense mechanisms per system asset.

\section{System Decomposition and Agile Threat Modeling based on 4+1 View Model Analysis} 
\label{section3}

This section introduces the \ac{iov} location service application as the base scenario for aligning assets with reference models.
Next, our 4+1 view model analysis in Sections \ref{section3lv} to \ref{section3phv} identify all sub-architectures as well as security domains of hardware, networking, software, and data.
This step marks one of our contributions and allows fine-grained identification and mappings of assets for our security verification method.

\subsection{Location Services of Connected Autonomous Vehicles} 
\label{scenariosec}

\begin{figure}[t]
    \centering
    \begin{tikzpicture}[scale=1]

    \draw [<->] (1.65,-0.5) -- ++ (0,-0.2) -- ++(2.5,0) -- ++(0,0.2);
    \draw [<->] (1.35,-0.5) -- ++ (0,-0.2) -- ++(-2.5,0) -- ++(0,0.2);
    \draw [<->] (1.65,3) -- ++ (0,0.2) -- ++(2.5,0) -- ++(0,-0.2);
    \draw [<->] (1.35,3) -- ++ (0,0.2) -- ++(-2.5,0) -- ++(0,-0.2);
    \draw [<->] (5.35,0.25) -- ++ (0.2,0) -- ++(0,1.75) -- ++(-0.2,0);
    \draw [<->] (-2.8,0.25) -- ++ (-0.2,0) -- ++(0,1.75) -- ++(0.2,0);
    \draw [<->] (1.5,-0.5) -- ++(0,-0.4) -- ++(-4.7,0) -- ++(0,3.1) -- ++(0.4,0);

    \draw pic[scale=0.2,transform shape = true] () at (1,0) {mybus={yellow}{blue!10}{black!60!green}};
    \draw pic[scale=0.2] () at (1,1) {antenna};
    \path (0.55,0.4) edge [decoration={expanding waves,angle=60,segment length=2mm},decorate,draw,color=black] (-0.2,0.4);
    \path (2.25,0.4) edge [decoration={expanding waves,angle=60,segment length=2mm},decorate,draw,color=black] (3,0.4);

    \draw pic[scale=0.75] () at (0,1.5) {basestation};
    \node[text width=2cm] at (2,2) 
    {Stationary GPS Receivers};
    \draw pic[scale=0.3,transform shape = true] (server1) at (3.35,1.6) {server={black!30!white}{white!30!blue}{white!60!blue}};
    \node[text width=2cm] at (5.2,2) 
    {A-GPS Server};
    
    \draw pic[scale=0.3,transform shape = true] () at (-2.25,1) {cellularsite};
    \node[text width=2cm] at (-0.75,0.35) 
    {Stationary DGPS};
    
    \draw pic[scale=0.3,transform shape = true] () at (3.6,1) {cellularsite};
    \node[text width=1cm] at (4.7,0.35) 
    {Base Station};
    
    \draw pic[scale=0.15,transform shape = true] () at (-2.4,2) {satelite={white!30!gray}{}};
    \node[text width=2cm] at (-0.75,2) 
    {Satellite};

    \end{tikzpicture}
  \caption{High-level application overview of the \ac{cav} location service scenario which includes A-GPS \cite{rubino2009gps}, DGPS \cite{pepe2018cors}, and Cellular Internet \cite{lee2014design} services.}
\label{fig:physicallocation}
\end{figure}
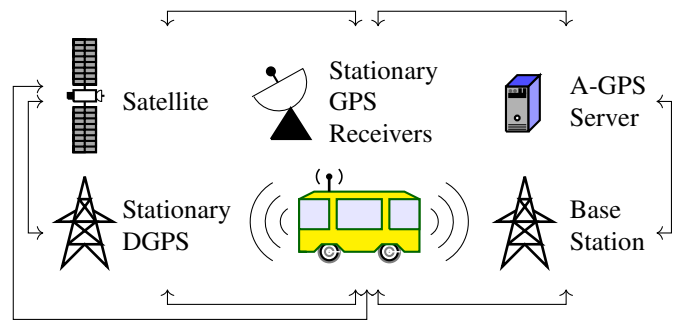
The location service application of \ac{cav}s represents the main scenario due to the following reasons.
Location accuracy contributes to the safety criticality level of a vehicle which, in turn, determines the level of driving automation of \ac{cav}s \cite{bolle2017early}.
Autonomous mini-bus shuttles of \ac{oem} aim towards running on \ac{sae} driving automation level four \cite{stocker2018shared}.
Driving automation level four expects automated steering, acceleration, deceleration, monitoring, handling of dynamic tasking, and driving modes of the vehicle system.
To prevent vehicles from stopping due to location inaccuracy, which causes a high safety criticality level, vehicles rely on redundant location services \cite{kolb2020technische}.

Several processing services of odometry, \ac{lidar}, and \ac{dgps} data, as shown in Figure \ref{fig:physicallocation}, establish necessary localization redundancy. 
Additionally, the comparison of calculations of local positions of sensor and receiver data with predetermined \ac{slam} trajectories enhances location estimation.
All in all, the services of vehicle \ac{dgps} communication, sensor (odometry, \ac{lidar}, camera) data processing and communication, \ac{slam} trajectory comparison, and \ac{v2c} communication make up the foundation of the following view model analysis.

With the scenario defined, it is necessary to determine the reference model domains of the \ac{iov} architecture for the analysis of the attack surface.
The work \cite{maple2019connected} provides a comprehensive \ac{iov} reference architecture which considers a physical \ac{iov} infrastructure consisting of four sub-architectures of \ac{cav}s, devices and peripherals, edge, and cloud. 
Their reference architecture for attack surface analysis is based on a functional-communication viewpoint which creates feasible complexity and manages incorporation of security relevant details.
To further simplify the sub-architecture categorization, we consider the peripherals and the vehicle as one domain. The reasons of (1) dynamic connectivity requirements that apply to vehicles and peripherals in the same way \cite{cheng2015routing} and (2) wired connections of peripherals to the in-vehicle network \cite{sharma2019extended} justify this assumption.

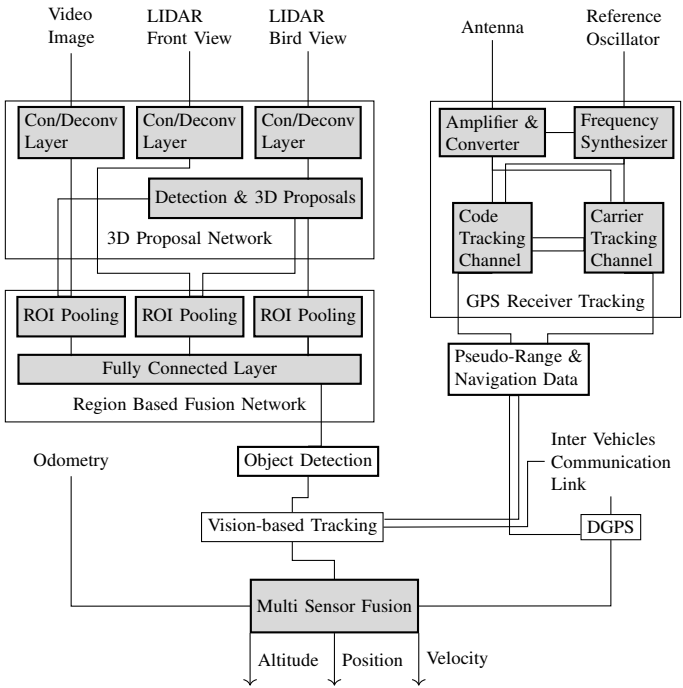
\begin{figure}[t]
    \centering
    \begin{tikzpicture}[scale=0.7]
    \node (ref1) at (0,6) [align=left, transform shape = true] {Video \\Image};
    \node (ref2) at (2.25,6) [align=left, transform shape = true] {LIDAR \\Front View};
    \node (ref3) at (4.5,6) [align=left, transform shape = true] {LIDAR \\Bird View};
    \node (netw1) at (2.25,3.1) [draw, align=left, minimum height=3cm,minimum width=7cm, transform shape = true] {\\ \\ \\ \\ \\ \\ 3D Proposal Network};
    \node (ac1) at (0,4) [rectangle, thick, draw=black, fill=white!70!gray, minimum width=1cm, minimum height=0.75cm, transform shape = true, align=left] {Con/Deconv \\ Layer};
    \node (ac2) at (2.25,4) [rectangle, thick, draw=black, fill=white!70!gray, minimum width=1cm, minimum height=0.75cm, transform shape = true, align=left] {Con/Deconv \\ Layer};
    \node (ac3) at (4.5,4) [rectangle, thick, draw=black, fill=white!70!gray, minimum width=1cm, minimum height=0.75cm, transform shape = true, align=left] {Con/Deconv \\ Layer};
    \node (pool1) at (0,0.5) [rectangle, thick, draw=black, fill=white!70!gray, minimum width=1cm, minimum height=0.75cm, transform shape = true, align=left] {ROI Pooling};
    \node (pool2) at (2.25,0.5) [rectangle, thick, draw=black, fill=white!70!gray, minimum width=1cm, minimum height=0.75cm, transform shape = true, align=left] {ROI Pooling};
    \node (mid1) at (3.5,2.75) [rectangle, thick, draw=black, fill=white!70!gray, minimum width=1cm, minimum height=0.75cm, transform shape = true, align=left] {Detection \& 3D Proposals};
    \node (pool3) at (4.5,0.5) [rectangle, thick, draw=black, fill=white!70!gray, minimum width=1cm, minimum height=0.75cm, transform shape = true, align=left] {ROI Pooling};
    \node (ac) at (2.25,-0.5) [rectangle, thick, draw=black, fill=white!70!gray, minimum width=6.5cm, minimum height=0.5cm, transform shape = true, align=left] {Fully Connected Layer};
    \node (netw1) at (2.25,-0.25) [draw, align=left, minimum height=2.5cm,minimum width=7cm, transform shape = true] {\\ \\ \\ \\ \\ Region Based Fusion Network};
    \node (ref5) at (4.5,-2.25) [draw, thick, align=left, transform shape = true] {Object Detection};
    \node (ref6) at (0,-2.25) [align=left, transform shape = true] {Odometry};
    \draw [-] (ref1.south) -- (ac1.north);
    \draw [-] (ref2.south) -- (ac2.north);
    \draw [-] (ref3.south) -- (ac3.north);
    \draw [-] (ac1.south) -- (pool1.north);
    \draw [-] (ac2.south) -- ++(0, -0.175) -- ++(-1.75,0) -- ++(0,-2) --++(1.75,0) -- (pool2.north);
    \draw [-] (ac3.south) -- ++(0,-0.4);
    \draw [-] (pool1.north) -- ++(-0.25,0) -- ++(0,1.85) -- ++(1.75,0);
    \draw [-] (pool2.north) -- ++(0.25,0) -- ++(0,0.435) -- ++(1.75,0) -- ++(0,1.05);
    \draw [-] (pool3.north) -- ++(0,1.5);
    \draw [-] (pool1.south) -- ++(0,-0.36);
    \draw [-] (pool2.south) -- ++(0,-0.36);
    \draw [-] (pool3.south) -- ++(0,-0.36);
    \draw [-] (ref5.north) -- ++(0.25, 0) -- ++(0,1.2);
    \node (netw4) at (9.2,2.54) [draw, align=left, minimum height=4cm,minimum width=4.75cm, transform shape = true] {\\ \\ \\ \\ \\ \\ \\ \\ \\ GPS Receiver Tracking};
    \node (visio) at (4.2,-3.5) [draw, align=left, transform shape = true] {Vision-based Tracking};
    \node (intev) at (10.25,-2.25) [align=left, transform shape = true] {Inter Vehicles \\Communication \\Link};
    \node (dgps) at (10.25,-3.5) [draw, align=left, transform shape = true] {DGPS};
    \node (seudo) at (8.5,-0.5) [draw, thick, align=left, transform shape = true] {Pseudo-Range \&  \\Navigation Data};
    \draw [-] (intev.south) -- (dgps.north);
    \draw [-] (intev.west) -- ++(-0.325,0) -- ++(0,-1.24) -- (visio.east);
    \draw [-] (pool3.south) -- ++(0,-0.36);
    \draw [-] (ref5.south) -- ++(0,-0.4) -- ++(-0.3,0) -- (visio.north);
    \node (ant) at (8,6) [align=left, transform shape = true] {Antenna};
    \node (refoz) at (10.5,6) [align=left, transform shape = true] {Reference\\Oscillator};
    \node (ac7) at (8,4) [rectangle, thick, draw=black, fill=white!70!gray, minimum width=1cm, minimum height=0.75cm, transform shape = true, align=left] {Amplifier \& \\ Converter};
    \node (ac8) at (10.5,4) [rectangle, thick, draw=black, fill=white!70!gray, minimum width=1cm, minimum height=0.75cm, transform shape = true, align=left] {Frequency \\ Synthesizer};
    \node (catc) at (8,2) [rectangle, thick, draw=black, fill=white!70!gray, minimum width=1cm, minimum height=0.75cm, transform shape = true, align=left] {Code \\ Tracking \\ Channel};
    \node (cotc) at (10.5,2) [rectangle, thick, draw=black, fill=white!70!gray, minimum width=1cm, minimum height=0.75cm, transform shape = true, align=left] {Carrier \\ Tracking \\ Channel};
    \draw [-] (ant.south) -- (ac7.north);
    \draw [-] (refoz.south) -- (ac8.north);
    \draw [-] (ac7.east) -- (ac8.west);
    \draw [-] (ac7.south) -- (catc.north);
    \draw [-] (ac7.south)  -- ++(0,-0.25) -- ++(2.25,0) -- ++(0,-0.6);
    \draw [-] (ac8.south)  -- ++(0,-0.1) -- ++(-2.25,0) -- ++(0,-0.75);
    \draw [-] (ac8.south) -- (cotc.north);
    \draw [-] (cotc.west) -- (catc.east);
    \draw [-] (cotc.west) -- ++(0,-0.25) -- ++(-1,0);
    \draw [-] (seudo.south) -- ++(0,-2.35) -- ++(-2.55,0);
    \draw [-] (seudo.south) -- ++(-0.175,0) -- ++(0,-2.64) -- ++(1.35,0);
    \draw [-] (catc.south) -- ++(-0.65,0) -- ++(0,-1.15) -- ++(1,0) -- ++(0,-0.2);
    \draw [-] (cotc.south) -- ++(0.55,0) -- ++(0,-1.15) -- ++(-2,0) -- ++(0,-0.2);
    \node (msf) at (5,-5) [draw, thick, align=left, transform shape = true, fill=white!70!gray, minimum height=1cm] {Multi Sensor Fusion};
    \draw [->] (msf.south west) -- ++(0,-1) node [midway, transform shape = true, right=0.025cm] (txt20) {Altitude};
    \draw [->] (msf.south east) -- ++(0,-1) node [midway, transform shape = true, right=0.025cm] (txt21) {Velocity}; 
    \draw [->] (msf.south) -- ++(0,-1) node [midway, transform shape = true, right=0.025cm] (txt22) {Position};
    \draw [-] (dgps.south) -- ++(0,-1.25) -- (msf.east);
    \draw [-] (ref6.south) -- ++(0,-2.47) -- (msf.west);
    \draw [-] (visio.south) -- ++(0,-0.35) -- ++(0.8,0) -- (msf.north);
    \end{tikzpicture}
  \caption{Logical View of GPS Receiver Tracking System \cite{navstar1996user}, Sensor Data Processing \cite{chen2017multi}, and Vehicle Localization Components \cite{vetrella2016differential}.}
\label{fig:logviewlocation}
\end{figure}

\subsection{Logical View Analysis (Data Management)}
\label{section3lv}

\subsubsection{System Modeling}

The works \cite{10.1007/978-3-030-45096-0_18} and \cite{suhr2016sensor} provide a general overview of the logical software design. 
Figure \ref{fig:logviewlocation} combines three detailed logical views where the top part consists of a \ac{gps} receiver tracking and vision-based object detection system. 
The object detection sub-module processes \ac{lidar} front and bird view data as well as video images.
Latest \ac{ml} frameworks for visual data processing rely on proposal networks for preprocessing that feed fully connected fusion networks \cite{10.1007/978-3-030-45096-0_18}.
The \ac{gps} receiver tracking system estimates signal traveling times using code and carrier synchronization techniques in order to determine first pseudo ranges. 
In cases of unreliable \ac{gps} signal reception, location services need to rely on predictive \ac{dgps} carrier phase corrections \cite{indriyatmoko2008artificial}.

The lower part of the logical view indicates a multi sensor fusion system which processes the results of vision-based tracking, \ac{gps} tracking, \ac{dgps} correction, and odometry data to determine high precision altitude, position, and velocity values.
The combination of \ac{lidar} point cloud and odometry data allows Kalman filter estimation of particle motion between \ac{lidar} scans. 
To determine a high precision offset of localization, it is possible to match point clouds between the estimated live \ac{lidar} scans and predefined \ac{slam} maps \cite{xu20173d}.

\subsubsection{Threat \& Vulnerability Analysis}
\label{datathreatvulnerability}

To calculate the \ac{cvss} scores of the logical view, it is necessary to gather the threats on identified \emph{assets} of the logical view. 
Logical software design, which describes the basic structure of data relationships and, thereby, application logic, belongs to field of system design engineering of software \cite{teorey2011database}.
Hence, the vulnerability taxonomies of \cite{igure2008taxonomies}, the \ac{risos} \cite{abbott1976security}, and \ac{pa} project \cite{bisbey1978protection}, which describe software \ac{os} flaws, apply to any other system design challenges as well. 
The reason for this is that an \ac{os} requires holistic system design modeling.

All possible software data flaws, of the referenced collection, affect the identified \emph{assets} of Figure \ref{fig:logviewlocation}.
For instance, incomplete or inconsistent parameter validation, privileges, identification, authentication, authorization, serialization, or logic errors mark vulnerabilities of the logical view domain.
By violating the vulnerabilities of data, an attacker may perform one or multiple \ac{crud} operations.
In our use case, data manipulation of applications of location services represents the ultimate goal of an attacker.


\subsubsection{Risk Analysis \& Security Requirements}
\label{section3rasr}


With the help of the threat and vulnerability analysis of the logical view, we apply risk analysis with the determination of the \ac{cvss} metric in Table \ref{tab:cvssscores}, where higher vulnerability scores refer to more severe risks.
This table indicates our decisions on \ac{cvss} parameters that we derive in Section \ref{analysisdigest}.
Logical security requirements require consideration of best practices of security by design concepts.
Another requirement is the incorporation of procedures of incident detection and reaction. 
Our methodology provides the possibility to cover existing types of such defense concepts in form of backward transition probabilities.

\subsubsection{Security Considerations}
\label{section3sc}

Based on the outcomes of the logical view analysis, a tailored security design from the logical perspective first of all needs to minimize the number of components and functionality requirements which belong to security by design concepts \cite{bishop1995taxonomy}. 
Other preventive measures such as error handling, consistency of data over time, authentication, validation, modularity, exposure, etc. require consideration and need incorporation into the logical design of the application scenario \cite{igure2008taxonomies}.
For detective and reactive measures, the logical design must detect injections of logic bombs which would intentionally hide, delete, or start processes that affect application logic.

\subsection{Developer View Analysis (Software Management)}
\label{section3dv}

\subsubsection{System Modeling}

The software implementation of location services builds upon the software stack of \ac{autosar} Classic and \ac{autosar} Adaptive which structure libraries, dependencies, and program interactions \cite{furst2016autosar}. 
The classic version of \ac{autosar} applies to deeply embedded systems that focus on safety, real-time capability, and process determinism.  
By contrast, the adaptive platform targets high-end computing in the form of custom applications.

The classic \ac{autosar} software architecture divides into four main layers. 
On top of the microcontroller layer, which groups \ac{ecu} hardware, the basic software layer as well as the \ac{autosar} runtime environment abstract hardware functionality through software modules. The top-level application layer utilizes the runtime environment for software and application module communication \cite{warschofsky2009autosar}. 
Equal to the classic \ac{autosar} architecture, the adaptive \ac{autosar} software architecture builds the adaptive \ac{autosar} foundation on top of a virtual machine/hardware layer.
The adaptive \ac{autosar} foundation consists of \ac{apis} and services for the management of the \ac{os}, time, execution, communication, configuration, security, and monitoring.
This layer enables the \ac{ara} to expose these \ac{apis} to applications that run on top of \ac{ara} \cite{furst2016autosar}.

Regarding the software architecture of the cloud, infrastructure, and edge domains of the reference model, the works \cite{bertaux2015software} and \cite{zhang2017carstream} introduce recent software networking stacks and cloud software architecture stacks respectively. 
These software assets mark potential entry points for an attacker and we consider this investigation as future work.

\subsubsection{Threat \& Vulnerability Analysis}
Since the developer view is part of the software context, it is possible to consider the traditional software threats of the \ac{stride} model.
Additionally, the software vulnerability taxonomy of \cite{tsipenyuk2005seven} lists input validation and representation, states of \ac{apis}, timing, errors, code quality, encapsulation, and environment as flaws.
These software flaws clearly focus on the implementation and software library modules and do not consider weak spots of data and system design.
All the stated flaws apply to the analysis of modules of software architecture which the developer view identifies.

\subsubsection{Risk Analysis \& Security Requirements}

Regarding the security requirements of the software context, the security requirements authenticity, integrity, non-repudiability, confidentiality, availability, and authorization of the \ac{stride} model apply.
With the help of the asset analysis of the developer view and the software vulnerability taxonomy, it is possible to determine the \ac{cvss} parameters of the software implementation layer in Table \ref{tab:cvssscores}. 
This software \ac{cvss} score represents the software risk analysis for the \ac{iov} location service scenario.

\subsubsection{Security Considerations}

Tailored security design in the software domain of the developer view concerns safe development, implementation, verification, testing, deployment, and maintenance of services such as interoperability, dynamic and automated risk assessment, attack prediction and attribution, threat predictive analytics, monitoring, and detection intelligence, encrypted traffic analysis, forensic readiness, intrusion detection and prevention, and penetration testing \cite{rao2017probabilistic}-\cite{haas2017intrusion}. 
It is necessary to apply all stated security concepts for location service networking, sensor fusion algorithms, modules, and dependencies of the \ac{os} in use.

\subsection{Process View Analysis (Networking Protocols)}
\label{section3prv}

\subsubsection{System Modeling}

\begin{figure}[t]
    \centering
    \begin{tikzpicture}[scale=0.9]

    \node (el11) at (0,2) [rectangle, anchor=center, thick, draw=black, fill=white!90!gray, minimum width=7cm, minimum height=0.5cm, transform shape = true] {GNSS Data Reception (A-GPS, DGPS)};
    \node (el12) at (0,1) [rectangle, anchor=center, thick, draw=black, fill=white!90!gray, minimum width=7cm, minimum height=0.5cm, transform shape = true] {Vision-based Tracking};
    \node (el13) at (0,0) [rectangle, anchor=center, thick, draw=black, fill=white!90!gray, minimum width=7cm, minimum height=0.5cm, transform shape = true] {Odometry(IMU) Trajectory Estimation};
    \node (el14) at (0,-1) [rectangle, anchor=center, thick, draw=black, fill=white!90!gray, minimum width=7cm, minimum height=0.5cm, transform shape = true] {Visual Scan and SLAM Map Matching};
    \node (el15) at (0,-2) [rectangle, anchor=center, thick, draw=black, fill=white!90!gray, minimum width=7cm, minimum height=0.5cm, transform shape = true] {Location Data Communication};
    
    \draw [->] (el11.south) -- (el12.north);
    \draw [->] (el12.south) -- (el13.north);
    \draw [->] (el13.south) -- (el14.north);
    \draw [->] (el14.south) -- (el15.north);

    \end{tikzpicture}
  \caption{High-level Process View of Localization Service.}
\label{fig:locationdetprocess}
\end{figure}
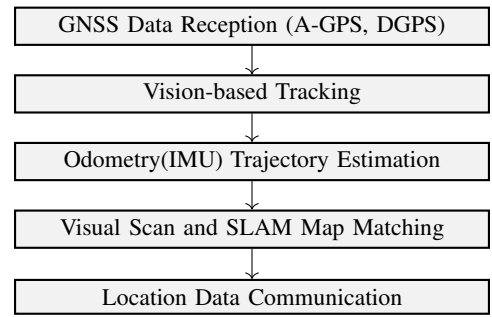

The process view indicates the interplay of logical components of localization services in a sequential order.  
Figure \ref{fig:locationdetprocess} represents the order which starts with \ac{agps} utilization for faster satellite localization. 
Reception of \ac{gps} data from satellites and subsequent merging of \ac{dgps} correction data determines the initial position estimation of the vehicle. 
At the same time, \ac{imu} data of vehicle movement passes the Kalman filter to feed the estimation between vision-based tracking scans. 
When it comes to the matching of scan and map data, the initial \ac{gps} position narrows the area of map matching which optimizes and stabilizes the position estimation \cite{qingqing2019multi}. 
The last part of the process view is the transmission of vehicle location data to the cloud services for vehicle tracking. 
Services of \ac{adas} modules such as maneuver estimation services \cite{alhajyaseen2012estimation} and lane hypothesis estimation benefit from the \ac{slam} map matching as well \cite{rabe2019lane}.

\subsubsection{Threat \& Vulnerability Analysis}
\label{section3prvtva}

All communication protocols of the mentioned services outside of the vehicle make up a direct attack surface for the attacker. 
Even though communication protocols differ, common attacks such as jamming, spoofing, timing, capturing, modification, removal, payload, etc. apply to all communication protocols independent of the software \ac{osi} reference layers \cite{cui2019review}.
The collection in \cite{igure2008taxonomies} provides network vulnerability taxonomies which equally apply in the \ac{iov} networking domain.






\subsubsection{Risk Analysis \& Security Requirements}
\label{section3prvrasr}

The risk analysis of the identified assets, threats, and vulnerabilities of the networking category provides another \ac{cvss} score of Table \ref{tab:cvssscores}.
Due to the exposure of communication messages and interfaces, it is necessary to emphasize on the reaction patterns of security requirements for the networking domain.
The large scale communication attack mitigation analysis of \cite{kholidy2016risk} identifies sixteen reactive defense mechanism and provides pros and cons of each mitigation strategy. 
Packet dropping, replication, isolation, disconnection, termination, restart, redirection, inspection, filtering, etc. belong to this collection of concepts. 
The defense mechanisms, thereby, counteract the malicious communicator and attacker types of sensor disruptor of the \ac{iov} specific attacker model of \cite{monteuuis2018attacker}.

\subsubsection{Security Considerations}

Security considerations for the networking domain affect the safe and reliable connectivity of \ac{v2i} and \ac{p2i}.
Since \ac{mitm} and other networking attacks are difficult to prevent, the focus in this domain lies on detection and, especially, reaction concepts \cite{khader2015preventing}.
Another reason for this fact is the necessary exposure of networking interfaces which enable the localization services in the first place. 
Hence, safe routing, redirect adaptivity, redundant connectivity, etc. point out the direction of tailored communication security for the location services of \ac{cavs}.






\subsection{Physical View Analysis (Hardware Management)}
\label{section3phv}

\subsubsection{System Modeling}

\begin{figure}[t]
    \centering
    \begin{tikzpicture}[scale=0.9]
    \node (g1) at (0,4) [rectangle, thick, draw=black, fill=white!80!gray, minimum width=1cm, minimum height=0.75cm, transform shape = true, align=left] {Gateway};
    \node (dc1) at (0.1,3) [rectangle, thick, draw=black, fill=white!70!gray, minimum width=1cm, minimum height=0.5cm, transform shape = true, anchor=west, align=left] {Domain \\Controller};
    \node (dc2) at (-0.1,3) [rectangle, thick, draw=black, fill=white!70!gray, minimum width=1cm, minimum height=0.5cm, transform shape = true, anchor=east, align=left] {Domain \\Controller};
    \node (points) at (-3,3) [transform shape = true, anchor=west] {.....};
    \node (points) at (4.2,3) [transform shape = true, anchor=west] {.....};
    \node (dc4) at (3.7,3) [rectangle, thick, draw=black, fill=white!70!gray, minimum width=1cm, minimum height=0.5cm, transform shape = true, anchor=east, align=left] {Domain \\Controller};
    \node (ec1) at (1.5,2) [rectangle, thick, draw=black, fill=white!90!gray, minimum width=1cm, minimum height=0.5cm, transform shape = true, anchor=west] {Sensing \& Diagnostics};
    \node (ec2) at (1.5,1) [rectangle, thick, draw=black, fill=white!90!gray, minimum width=1cm, minimum height=0.5cm, transform shape = true, anchor=west, align=left] {Adaptive Cruise \\Control Module};
    \node (ec4) at (-1.6,1) [rectangle, thick, draw=black, fill=white!90!gray, minimum width=1cm, minimum height=0.5cm, transform shape = true, anchor=east, align=left] {Telematics \\Module};
    \draw [-, thick] (g1.west) -- ++(-0.2,0) -- (dc2.north);
    \draw [-, thick] (g1.east) -- ++(0.2,0) -- (dc1.north);
    \draw [-, thick] (g1.east) -- ++(2.1,0) -- (dc4.north);
    \draw [-, thick] (dc1.south) -- ++(0,-1.535) -- (ec2.west);
    \draw [-, thick] (dc4.south) -- ++(0,-0.25);
    \draw [-, thick] (dc2.south) -- ++(0,-1.535) -- (ec4.east);
    \end{tikzpicture}
  \caption{Physical View of Simplified In-Vehicle E/E-Architecture}
\label{fig:physicalautomotivearch}
\end{figure}
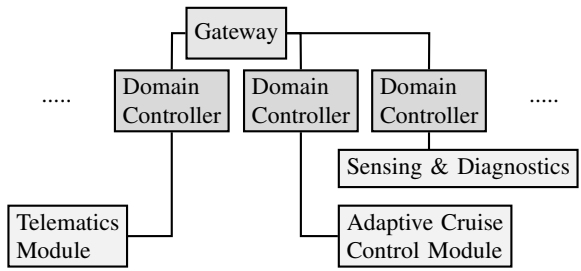

Figure \ref{fig:physicalautomotivearch} presents a simplified physical view of the in-vehicle architecture. 
There exist different architecture designs such as zone, domain, or central gateway based architectures \cite{brunner2017automotive}. 
The reference architecture shown in Figure \ref{fig:physicalautomotivearch} follows the domain-based architectural design. 
The reason for it is that the modern anatomy of automotive Ethernet has computationally powerful domain controllers which group \ac{adas}, drive-train, infotainment, \ac{hmi}, etc. network segments \cite{sharma2019extended}. 
The design enables isolation, criticality, and bandwidth measures to unload the gateway component \cite{shreejith2017vega}.

The automotive Electrical/Electronic-Architecture attaches sensors and actuators to \ac{ecu}s which in turn connect to domain controllers or directly connect to the central gateway component depending on safety critical functionality \cite{shreejith2017vega}. 
With the transmission of location data to the cloud, the gateway enables cloud services to publish vehicle information to smartphone applications \cite{lee2014design}.
Our physical analysis neglects the focus on infrastructure for \ac{slam} map construction as it happens before \ac{cav} deployment \cite{li2016road}.



\subsubsection{Threat \& Vulnerability Analysis}
\label{section3phytva}

It is unlikely for an attacker to gain physical access to infrastructure units in the cloud, networking, or satellite domain due to their remote location. 
For this reason, we focus on the threats and vulnerabilities of the physical vehicle architecture.
The general attack taxonomy of physical attack on \ac{iot} devices of \cite{abdul2018comprehensive} counts twelve types of hardware threats.
Here, threats and attacks map to affected security requirements and countermeasures. 
Object tampering, outage, object replication, camouflage, side-channel, hardware trojans, physical damage etc. attacks are among the threats of the work in \cite{abdul2018comprehensive}.

Highlighting in-vehicle attacks specifically, the work in \cite{sharma2019extended} provides a detailed attack surface.
Here, non-\ac{can} attacks, in the form of \ac{tpms} and KeeLoq Cipher, and \ac{can} attacks, in the form of media player, \ac{obd}, bluetooth module, and \ac{tcm} attacks, exploit physical vulnerabilities of the listed devices.
The vulnerability assessment of \cite{salfer2015attack} further identifies boot memory, debug interfaces, inter-chip communication channels, and side-channel attacks as susceptible hardware units.


\subsubsection{Risk Analysis \& Security Requirements}
\label{section3phyrasr}

With the attack surface, threat modeling, and vulnerability analysis, it is possible to calculate the \ac{cvss} scores of physical assets in Table \ref{tab:cvssscores}.
Regarding physical security requirements, the taxonomies in \cite{sharma2019extended} mention monitoring for intrusion detection as well as authentication as the main requirements.
To further protect hardware vulnerabilities, \cite{salfer2015attack} and \cite{oyler2016security} emphasize on stack canaries, no execute bit, address space layout randomization, protection units, management units, privilege separation, and \ac{hsm} mitigation concepts.

\subsubsection{Security Considerations}
\label{section3physc}

Opposed to networking components, access to physical components remains a challenging task due to location and speed dynamics of vehicles and the distance to cloud or infrastructure assets. 
Thus, tailored security for physical components of the \ac{iov} location service focuses on insider attacks \cite{dietzel2015context}.
This means physical attack surfaces such as \ac{obd} assets require misbehavior detection frameworks and secure aggregation mechanisms.


\section{Vulnerability Scores and Markov Chain-based Security Verification}
\label{section4}

This section walks through each \ac{cvss} vulnerability metric and defines each metric per view model perspective. 
All abbreviations used throughout this section refer to \ac{cvss} parameters and can be found in Table \ref{tab:cvssscores}.
Our scores mark the first input to probability calculations for state transition of our Markov Chain model.
Section \ref{evalmodel} describes our quantitative system-level security verification concept.
The second input for our Markov Chain model are attack vectors that contain assets for the system-level security verification. Possible attack vectors are presented in the evaluation Section \ref{section4avc}.

\subsection{Labeling of CVSS Parameters}
\label{analysisdigest}


The connectivity of the \ac{iov} architecture components enables the label "remote" (R) for the \ac{av} in every category. 
The \ac{ac} has a similar distribution where every category except networking fulfills the label "high" (H). 
The reason for this choice is the safety-critical application of \ac{cav} which requires the highest access control standards at every stage.
Networking \ac{ac} remains "low" (L) in the location service scenario due to the fact that attackers face direct access to networking applications of redundant location services.

Regarding authentication (A), software provides data access and authentication privileges per default. 
To access the \ac{iov} cloud and vehicle environment "requires" (R) authentication but infrastructure services such as \ac{gps} data reception does "not require" (N) authentication.
Every category requires authentication concepts except the data domain. 
Regular \ac{gps} receivers do not necessarily authenticate satellites. 
However, software behind signal reception interfaces authenticates correct signals.

Compromising software has the potential to cause "complete" (C) confidentiality, integrity, and availability loss in the system.
Equally, successful data and networking integrity manipulation could allow data or network participants to propagate through the system, if not correctly detected in initial checks. 
Otherwise, the impact on the confidentiality, integrity, and availability requirements remains "partial" (P).

\begin{table}[t] 
\renewcommand{\arraystretch}{1.3}
\caption{\ac{cvss} Scores per View Model Layer}
\label{tab:cvssscores}
\centering
\begin{tabular}{L{2.25cm}R{0.75cm}R{1cm}R{1.5cm}R{1.25cm}}
\hline
\textbf{Parameters} & \textbf{Data} & \textbf{Software} & \textbf{Networking} & \textbf{Hardware} \\ 
\hline
\hline
Access Vector  & R  & R  & R  & R \\ 
\hline
Access Complexity & H  & H  & L & H \\ 
\hline
Authentication & N  & R  & R  &  R \\ 
\hline
Confidentiality Impact & P  & C & P & P \\  
\hline
Integrity Impact & C & C & C & P \\ 
\hline
Availability Impact & P & C & P & P \\ 
\hline
Impact Bias & I & A & I & N \\ 
\hline
\textbf{Base Score} & \textbf{6.8}  & \textbf{4.8}  & \textbf{5.1}  & \textbf{3.4} \\ 
\hline
\hline
Exploitability & PoC & U & F & PoC \\ 
\hline
Remediation Level & TF & OF & TF & OF \\  
\hline
Report Confidence & UCB & UCF & UCB & UCF \\ 
\hline
\textbf{Temporal Score} & \textbf{5.2}  & \textbf{3.2}  & \textbf{4.1}  & \textbf{2.4} \\ 
\hline
\hline
Collateral Damage Potential & H & H & M & M \\ 
\hline
Target Distribution & M & L & H & L \\ 
\hline
\textbf{Environmental Score} & \textbf{5.7} &  \textbf{1.6} & \textbf{5.3}  & \textbf{1.2} \\ 
\hline
\hline
\textbf{Total} & \textbf{17.7} &  \textbf{9.6} & \textbf{14.5}  & \textbf{7} \\ 
\hline
\end{tabular}
\end{table}

For the \ac{ib} and with regard to the location service scenario, data and networking components weight "integrity" (I) over other requirements, as incorrect location data or communication entities potentially destroy the service.
Since there is a centralized sensor fusion software module, the \ac{ib} of software applies greater weighting to "availability" (A).
With respect to the hardware category, exploiting any of the listed security requirements leads to comparable "normal" (N) impact of the attack on the system.
Regarding data attacks, existing research on \ac{gps} spoofing provide a "proof of concept" (PoC) to manipulate location data \cite{larcom2013modeling}. 
This fact can be used to assume the existence of additional "uncorroborated" (UCB) sources for the report confidence (RC).
At the same time and concerning the Remediation Level (RL), "temporal fix" (TF) solutions exist for the detection and prevention of such attacks.


The networking category behaves similar except that it is possible to access "functional" (F) exploit code for networking attacks by using specific \ac{os}s for hacking.
With software, it is possible to assume non-disclosed algorithms which implement sensor fusion and localization. 
This fact sets the exploitability (E) of location service \ac{ecu} software to "unproven" (U).
Due to the criticality of location service correctness, one must expect "official fixes" (OF) of newly confirmed vulnerabilities.
However, if software bugs remain undiscovered, they remain "unconfirmed" (UCF) from the report confidence perspective.

The collateral damage potential (CDP) of data and software is "high" (H), as it directly affects system safety.
Redundancy and robustness of location services enables 
temporal autonomy of a vehicle and reduces the damage potential of networking and hardware attacks to "medium" (M) \cite{miucic2018connected}.
Regarding target distribution, the multi sensor fusion software as well as the physically reachable hardware deserve a "low" (L) target distribution (TD) value. 
Communication and location data propagate from infrastructure nodes through the vehicle to the cloud and require a "high" (H) distribution value. 
However, compromised location data of cloud services does not affect the location service functionality of the vehicle itself.
Only the redistribution of malicious location data from cloud services to other vehicles causes problems. 
For this reason, the evaluation labels the distribution of highly critical location data as "medium" (M).

With all parameters specified, it is possible to calculate the overall \ac{cvss} scores \ac{bs}, \ac{ts}, and \ac{es}. The equations \ref{cvssf1}, \ref{cvssf2}, and \ref{cvssf3} calculate the main \ac{cvss} scores and can be found in \cite{schiffman2004common}, where the values of \ac{cib}, \ac{iib}, and \ac{aib} depend on the setting of the \ac{ib}.
\begin{equation}
\label{cvssf1}
\resizebox{0.91\hsize}{!}{$
BS = 10 \cdot AV \cdot AC \cdot A \cdot ((CI \cdot CIB)+(II \cdot IIB)+(AI \cdot AIB))
$}
\end{equation}
\begin{equation}
\label{cvssf2}
\resizebox{.35\hsize}{!}{$
TS = BS \cdot E \cdot RL \cdot RC
$}
\end{equation}
\begin{equation}
\label{cvssf3}
\resizebox{.54\hsize}{!}{$
ES = (TS+(10-TS) \cdot CDP) \cdot TD
$}
\end{equation}

\subsection{Quantitative Security Verification Model}
\label{evalmodel}

It is possible to choose a slightly simplified version of the attack realization metric and algorithm in \cite{cheng2012research} to demonstrate the applicability of extended Markov Chain models on attack propagation graphs that follow the categorization structures of the 4+1 view model analysis. 
The reason not to rely on non-homogenous continuous-time Markov models, as in \cite{abraham2016estimating}, is the fact that the extended Markov Chain suffices in modeling the high-level attack stages of our \ac{iov} use cases.


The discrete-time finite state Markov Chain represents a time and state discrete stochastic process where future states at time $t_{i+1}$ depend on current states at time $t_{i}$ only, without relying on past states at time $t_{i-1}$.
Per definition, the Markov Chain $MC(I, P, A)$ is a 3-tuple consisting of system state space $I$, transition probability matrix $P$, and a set of possible atomic actions $A$.
We assume no empty action that affects the realization metric $E$, hence, setting it to $E=1$.
To further simplify, it is possible to remove both sums of the state to target probability.
The reason for this is the interest in the worst-case attack with maximum significance.
This fact maintains state transitions that connect starting and target states without detours.
As a result, the following characteristics count for (1) state, (2) transition, (3) action, and (4) total state to target probability of attack realization respectively:
\begin{enumerate}
    \item $S_{i} \in I$, where the state $S_i$ has one of the labels of $HW$, $SW$, $Net$, or $Data$ of the view model perspectives. 
    \item $\sum_{j=1}^{\infty} p_{ij} = 1$, $\forall p \in P$. %
    \item $a_{i}, d_{i} \in A$ are probabilities of successful attacks and defense mechanisms. 
    \item $W^{n}(S_{i=1})= \sum_{S_i \in \text{SUBSEQ}(S_1)} p_{ij} \cdot W^{n-1}(S_i)$, where SUBSEQ returns the set of remaining states $S_i$. 
\end{enumerate}


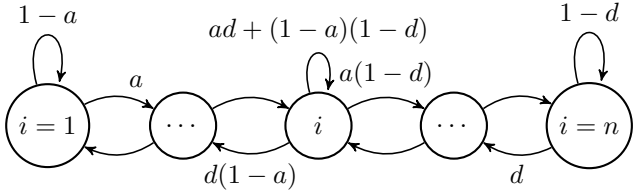
\begin{figure}[!ht]
    \centering
    
    \begin{tikzpicture}[scale=0.9, ->, >=stealth', auto, semithick, node distance=1.8cm]
        \tikzstyle{every state}=[fill=white,draw=black,thick,text=black,scale=1]
            \node[state]    (A)                     {$i=1$};
            \node[state]    (B)[right of=A]   {$\dots$};
            \node[state]    (C)[right of=B]   {$i$};
            \node[state]    (D)[right of=C]   {$\dots$};
            \node[state]    (E)[right of=D]   {$i=n$};
            \path
            (A) edge[loop above]     node{$1-a$}         (A)
                edge[bend left]     node{$a$}     (B)
            (B) edge[bend left]    node{}           (C)
                edge[bend left, below]    node{}           (A)
            (C) edge[bend left]    node{$a(1-d)$}   (D)
                edge[loop above]     node{$ad+(1-a)(1-d)$}   (C)
                edge[bend left, below]     node{$d(1-a)$}   (B)
            (D) edge[bend left]     node{}     (E)
                edge[bend left, below]    node{}    (C)
            (E) edge[loop above]    node{$1-d$}     (E)
                edge[bend left,below]    node{$d$}      (D);
        \end{tikzpicture}
    
  \caption{Markov Chain Transition Probability Graph of Attack and Reaction Transition Probabilities}
\label{fig:transitionprobgraph}
\end{figure}



The state transition probabilities, shown in Figure \ref{fig:transitionprobgraph}, include attack $a$ and defense $d$ actions. 
Only the initial state starts with either a successful attack or remains in the set of initial states. 
Similarly, the last state changes through an effective response action against the attacker only.
For all intermediate steps, there is no state transition if a successful attack faces an immediate countermeasure (ad), or neither an attack nor a defense action happens $((1-a)(1-d))$.
If an attack action succeeds and no defense reaction occurs $a(1-d)$, the attacker moves to the next state.
Vice versa, failing attacks and successful reactions $d(1-a)$ may transition the attacker backwards.
Section \ref{evalassumptions} provides sample calculations of the probabilities which enable the quantitative security verification.
As a last rule, outgoing weights of each state sum up to the value of one. 
\begin{equation}
\label{formula0}
a = \frac{f_{cvss} (v_{\text{domain}})}{42.5}; \hspace{5mm} i \in \mathbb{N}.
\end{equation}
\begin{equation}
\label{formula02}
a_i = 1-e^{-2i\frac{f_{cvss} (v_{\text{domain}})}{42.5}}; \hspace{5mm} i \in \mathbb{N}.
\end{equation}

It is possible to model parameters of the attack and defense probabilities with the \ac{cvss} vulnerability assessment. 
Additionally, the stage of the Markov Chain depends on the view model perspectives of type hardware, networking, software, and data of the asset under attack.
Since the \ac{cvss} score has a maximum possible value of $42.5$ (choose largest possible value for every \ac{cvss} parameter), it is possible to normalize each vulnerability score with respect to this value.
Equation \ref{formula0} shows the resulting attack probability, where $i$ refers to the attacking stage of the attack vector.
The attack stage determines what view model perspective type to choose.
It is possible to improve the model of the attack probability $a$ per stage $i$ by introducing Equation \ref{formula02} (adapted from the work in \cite{mcqueen2006time}).
Equation \ref{formula02} describes an increase of the attacking likelihood for increasing stages $i$.
The underlying assumption of Equation \ref{formula02} is the fact that a single successful attack opens up opportunities to compromise more vulnerabilities or combinations of vulnerabilities. 
More chances for the attacker to find vulnerabilities increases the likelihood of a successful attack.

The defense mechanism probabilities $d$ depend on actual attack actions. 
It is possible to model probabilities of successful countermeasures independent of the attack due to missing attack attribution formulas which would enable attack identification, assessment, and reaction actions for all attacking stages.
This measure simplifies the quantitative calculations in the Markov Chain model, but requires further investigation in the future.

\begin{table*}[t]
\renewcommand{\arraystretch}{1.3}
\caption{IoV Cyber Attack Path Propagation}
\label{tab:attackpropagation}
\centering
\begin{tabular}{L{1em}L{8em}R{10em}L{42em}}
    \hline
    \textbf{ID}& \textbf{Attacker Type} & \textbf{Model Type} & \textbf{Sample Attack Path Propagation}  \\
    \hline
    \hline
    1 & Unauthorized & Cloud & Browser redirect attack \& Shell access (C-Net) $\Rightarrow$ Privilege escalation (C-SW) $\Rightarrow$ Access to ECU (V-Net) $\Rightarrow$ CAN bus attack (V-Data) \cite{maple2019connected} \\ 
    \hline
    2 & Unauthorized & Infra \& Edge & Road sign attack (I-HW (a) or I-Net (b)) $\Rightarrow$ Road sign distortion (I-Data) $\Rightarrow$ Camera image data modification (V-Data) \cite{hamad2020savta} \\ 
    \hline
    3 & Unauthorized &  Vehicle \& Peripherals & Eavesdropping wireless \ac{tpms} (V-Net) $\Rightarrow$ Reverse engineering attack (V-SW) $\Rightarrow$ Packet injection attack (V-Data) \cite{rouf2010security} \\ 
    \hline
    4 & Authorized & Cloud \textbf{or} Infra \& Edge & Malicious software update (V-SW) \& Driver assistance attack (V-Data) \cite{steger2016secup} \\ 
    \hline
    5 & Authorized & Vehicle \& Peripherals & Disabled ECU hardening \& CAN replay attack (V-Data) \cite{xiong2019threat} (based on \cite{miller2015remote}) \\ 
    \hline
    
  \end{tabular}
\end{table*}

\section{Evaluation of Quantitative system-level security verification}
\label{section5}

This section performs and evaluates our quantitative system-level security verification on a selection of attack paths that contain assets of the location service scenario. To do so, Section \ref{section4avc} introduces chosen attack vectors. 
The assets of these attack paths are allocated to 4+1 view perspectives.
Section \ref{evalassumptions} utilizes the assets of the attack vectors and applies them together with our \ac{cvss} scores to the Markov Chain verification model.
The last section lists the results of the security verification and evaluates its features and trends.

\subsection{Selection of IoV Attack Paths}
\label{section4avc}

Attack vectors define the points of an infrastructure where an attacker enters the system unauthorized. The sum of an attack vector represents the attack surface which is what an attacker faces when attacking a system \cite{10.5555/3067087}. There are different methods for an attacker to enumerate, analyze, exploit, and enter the attack surface \cite{al2016cyber}. Afterwards, an attacker follows an arbitrary path until she reaches the target. Regarding attack propagation characteristics, attacking capability and scope of the attacker model remain the dominating properties for the determination of the depth of the attack \cite{de2014security}.

For the scope of our work, we consider unauthorized as well as authorized attackers with equal skill level to specify different initial starting points for attacks. Table \ref{tab:attackpropagation} shows our selection of \ac{iov} attacks that contain assets identified during the 4+1 view model analysis in Section \ref{section3}. The attack path of attack with ID 1 start in the cloud domain to eventually compromise the vehicle location service by provoking a lane departure. With the help of our analysis, the affected assets at different stages of the attack can be mapped to a networking attack in the cloud (C-Net), software compromise in the cloud (C-SW), in-vehicle network attack (V-Net), and vehicle data attack (V-Data). 
The reason for grouping the propagation stages with regard to hardware, networking, software, and data attacks serves for asset to view category mapping to facilitate the application of our Markov Model for security verification.

The infrastructure attack with ID 2 initially targets road signs to cause distortions in camera images that are processed by the 3D proposal network and thus, the multi sensor fusion unit. The attack with ID 3 directly targets the vehicle \ac{tpms} with the intention to either stop or compromise vehicle privacy (tracking location data) by indicating wrong tire pressure values. Analysing this variety of attacks with different length of attack paths allows to investigate the behavior of our security methodology as well as if our methodology can be applied to any attacks in the \ac{iov} infrastructure. 
The assumption for the authorized attack with ID 5 is a malicious but trusted developer with \ac{obd} and \ac{ecu} authentication credentials. For this type of attacker, performing CAN replay attacks to eventually affect vehicle trajectories of the \ac{adas} system should not be possible.

Future replacement of \ac{ecu} software modules in \ac{autosar} adaptive requires the update and configuration service within \ac{ara} to check integrity, authenticity, and sometimes confidentiality of module binaries \cite{steger2016secup}. We assume that attack path with ID 4 requires an authorized attacker with knowledge of security credentials (symmetric cryptography session keys as well as asymmetric cryptography key pairs) to pass integrity, authenticity, and confidentiality checks of the wireless communication service. Such an attack can be performed by stealing credentials from dedicated communication devices located in the cloud or \ac{iov} infrastructure. For the attack target, we assume an \ac{ecu} software module running a location service component (e.g. part of \ac{adas}).

\subsection{Evaluation of our Methodology (Perform Security Verification of Attack Paths)} 
\label{evalassumptions}



%
%

In order to evaluate the 4+1 view model analysis in the security context, we apply our methodology to our selection of attacks (see Table \ref{tab:attackpropagation}). With the 4+1 view model analysis, it becomes feasible to allocate every asset of \ac{iov} attack paths to one of the domains. At the same time, each view model domain marks a stage of our Markov Chain transition model to verify security quantitatively.
The following paragraphs demonstrate the process of applying one of the attack paths to our results gained from the 4+1 view model analysis. Afterwards, we contrast the results of different attack paths to determine the behavior, features, and possibilities of our concept. For showcasing the application of the view model security verification concept, we consider the attack path with ID 1, consisting of a could network (C-Net) and software (C-SW) attack as well as vehicle networking (V-Net) and data (V-Data) attack.

The values for the calculation of the Markov Chain state transition matrix $P$ depend on the probabilities of successful attacks and patches. 
Table \ref{tab:cvssscores} shows the vulnerability scores per asset for each domain of the view model perspectives. 
Higher values determine a higher likelihood of attacking an asset successfully.
The domain specific \ac{cvss} score over the maximum possible \ac{cvss} score determines the attacking probability $a$ (see Formula \ref{formula0}).
Considering the initial cloud networking attack of attack path with ID 1, the attacking probability $a$ depends on the networking stage \ac{cvss} score $f_{cvss}(\text{Net}) = 14.5$ and calculates as shown in Equation \ref{formula1}. To simplify the evaluation, we leverage a constant value for a successful countermeasure probability $d$, which can be seen in Equation \ref{formula2}.
\begin{equation}
\label{formula1}a=\frac{14.5}{42.5}=0.34; \hspace{1mm} (1-a)=\frac{28}{42.5}=0.65
\end{equation}
\begin{equation}
\label{formula2}
d=\frac{1}{10}=0.1; \hspace{1mm} (1-d)=\frac{9}{10}=0.9
\end{equation}

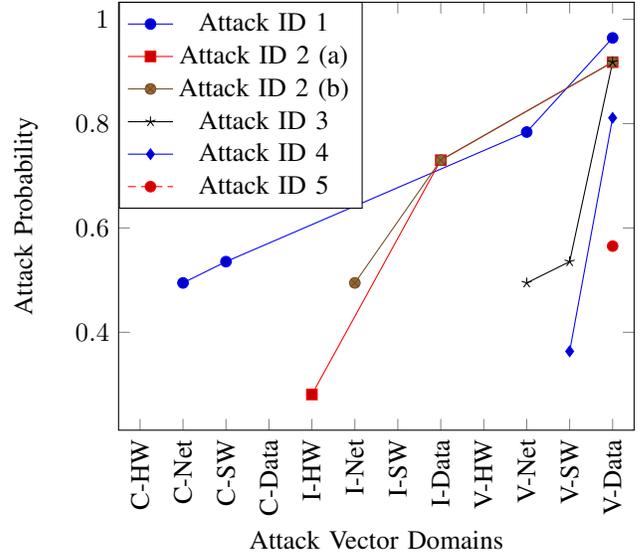
\begin{figure}[t]
\centering
\begin{tikzpicture}
\begin{axis}[scale=1,
    legend style={at={(0.237,1.001)},anchor=north},
    xmin=-0.5,xmax=11.5,xticklabels={C-HW,C-Net,C-SW,C-Data,I-HW,I-Net,I-SW,I-Data,V-HW,V-Net,V-SW,V-Data},xtick={0,...,11},x tick label style={rotate=90,anchor=east},
    x label style={at={(axis description cs:0.5,-0.12)},anchor=north},
    xlabel={Attack Vector Domains},
    ylabel={Attack Probability}
    ],
    \addplot table [y=ind1] {performance1.data};
    \addplot table [y=ind3] {performance1.data};
    \addplot table [y=ind4] {performance1.data};
    \addplot table [y=ind5] {performance1.data};
    \addplot table [y=ind2] {performance1.data};
    \addplot table [y=ind6] {performance1.data};
    
    \addlegendentry{Attack ID 1}
    \addlegendentry{Attack ID 2 (a)}
    \addlegendentry{Attack ID 2 (b)}
    \addlegendentry{Attack ID 3}
    \addlegendentry{Attack ID 4}
    \addlegendentry{Attack ID 5}
\end{axis}
\end{tikzpicture}
\caption{Attack probabilities (see Equation \ref{formula02}) of attack paths (Table \ref{tab:attackpropagation}) at different domain stages $i$ with \ac{cvss} score $f_{cvss}(i)$.} 
\label{fig:5}
\end{figure}

With the attack path with ID 1, the attack stages one to four consist of type C-NET, C-SW, V-Net, and V-Data. Furthermore, considering the attack forward transition probabilities $a_1 = a$ at stage $i=1$, $a_{i}=a(1-d)$, and $a_{i=n}=0$ of Figure \ref{fig:transitionprobgraph}, the attack probabilities calculate as follows: At stage $i=1$, the first attack transition probability calculates as $a_1= 1-e^{-2(\frac{14.5}{42.5})} = 0.4946$, assessing a could networking attack.
Subsequent stages calculate as $a_2=(1-e^{-4(\frac{9.6}{42.5})})\cdot(1-0.1) = 0.53541$, $a_3=(1-e^{-6(\frac{14.5}{42.5})})\cdot(1-0.1) = 0.78381$, and $a_4=(1-e^{-8(\frac{7}{42.5})})\cdot(1-0.1) = 0.9643$, assessing a cloud software, vehicle network and data attack respectively.
Figure \ref{fig:5} indicates these numbers with the blue line of the attack path with ID 1.
For a total attack probability (includes all forward transition probabilities), the product of these values result in $a_1*a_2*a_3*a_4=0.2001=20\%$ as indicated in Table \ref{tab:results}.
Other values of Table \ref{tab:results} correspond to all other attack paths of Table \ref{tab:attackpropagation}, where Figure \ref{fig:5} shows intermediate probability values.


Our results show that with the 4+1 view model analysis, arbitrary \ac{iov} attack paths can be mapped to view model domains. This enables comparative and quantitative system-level security verification of system assets. 
The attack realization probabilities of the initial states align with the expected behavior of lower probable attacks for longer attack paths.
The lower percentages for paths that originate from the cloud and infrastructure locations confirm this claim.

Furthermore, authorized attackers have higher probabilities to successfully attack localization services which aligns with expectations.
This fact can be seen when inspecting authorized versus unauthorized attack probability results. 
This outcome makes sense due to the possible size of an in-vehicle attack propagation compared to the Internet attack propagation path.
Similarly, the results of direct vehicle or close proximity attacks are more likely to affect location data. 
An explanation could be that the set of attacks of the local attacker contains the attacks of the remote attacker as subset.

It is important to emphasize that changing our assumptions and with that the \ac{cvss} parameterization changes the outcomes of the security verification. Hence, the variance in the security verification results depends on our \ac{iov} driven parameterization.
Additionally, the decomposition of multiple views into more fine-grained categories lowers the attack probabilities drastically (longer paths) due to the multiplicative aggregation. 
Here, additional calculation are required to stabilize the multiplications of numbers lower than one.
In general, it is possible to state that the level of detail should remain similar for security designs with comparable complexity.

\begin{table}[t] 
\renewcommand{\arraystretch}{1.3}
\caption{Attack Realization Probabilities of All Initial Attack States}
\label{tab:results}
\centering
\begin{tabular}{L{2cm}R{1.5cm}R{2cm}R{1.5cm}} 
\hline
\textbf{Attacker Type} & \textbf{Cloud} & \textbf{Infra} \& \textbf{Edge}  & \textbf{Vehicle}  \\  
\hline
\hline
Authorized & 29.47 \% & 29.47 \% & 56.52 \% \\ 
\hline
Unauthorized & 20.01 \% & 18.80 \% (a) 33.13 \% (b) &  24.30 \%  \\ 
\hline
\end{tabular}
\end{table}

\section{Conclusion}
\label{section7}

This paper applies the well-established 4+1 view model in the security context of the \ac{iov} and utilizes agile threat modeling and risk assessment for a structured identification and security assessment of \ac{iov} assets.
The view model analysis separates data, software, networking, and hardware categories and enables the allocation of attack path assets to these respective domains.

With the mapping of attack path assets to respective 4+1 view model domains, our Markov Chain model uses state transition probabilities to assess attack and defense probabilities of individual assets. Attack paths with comparable size allow system-level security verification of multiple \ac{iov} assets.
The results show the applicability of our methodology to arbitrary \ac{iov} assets included in attack paths. Our \ac{cvss} parameterization is driven by the \ac{iov} infrastructure analysis and indicates security critical parts of \ac{iov} architecture.

\subsection{Future Work}
\begin{itemize}
    \item To support the quantitative security verification results based on the 4+1 view model analysis, hacker teams need to conduct comprehensive and multidisciplinary methodologies such as QuERIES \cite{carin2008cybersecurity}. 
    
    
    \item No research has been conducted with regard to automation of the security analysis approach. 
    To cope with complex systems of the \ac{iov}, automation of analysis concepts is mandatory for system wide security coverage \cite{rao2017probabilistic}.
    Here, it is possible to utilize existing automated threat modeling and risk assessment tools, as in \cite{xiong2019threat}, on separated perspectives. 
    
    \item The \ac{cvss} is an old vulnerability scoring system which is not tailored to IoV specific properties. 
    By the time of writing this paper, the work \cite{lee2020practical} introduced a new vulnerability scoring system that is more tailored to cover vulnerabilities of the \ac{iov}.
    Changing our assumptions and parameterization of \ac{cvss} scores changes our security verification results.

\end{itemize}

\ifCLASSOPTIONcaptionsoff
  \newpage
\fi



\bibliographystyle{IEEEtran}
\bibliography{bibtex/IEEEabrv,bibtex/mybibtex}

\end{document}